# Reconfigurable integrated full-dimensional optical lattice generator

Shuang Zheng[1,2]†, Jing Du[1,2]†, Xiaoping Cao[1,2], Jinrun Zhang[1,2], Zhenyu Wan[1,2], Yize Liang[1,2], Jian Wang[1,2]*

[1] Wuhan National Laboratory for Optoelectronics and School of Optical and Electronic Information, Huazhong University of Science and Technology, Wuhan 430074, Hubei, China.

[2] Optics Valley Laboratory, Hubei, Wuhan 430074, China

† These authors contributed equally to this work.

*Corresponding author: jwang@hust.edu.cn

**Optical lattices with periodic potentials have attracted great attention in modern optics and photonics, enabling extensive applications in atomic manipulation, optical trapping, optical communications, imaging, sensing, etc. In the last decade, the generation of optical lattices has been widely investigated by various approaches such as multi-plane-wave interferometer, beam superposition, spatial light modulators, nanophotonic circuits, etc. However, all of the previous state-of-the-art works are restricted to only one or two dimensions of the light field, which cannot fulfill the increasing demand on complex light manipulation. Full-dimensional and dynamic control of the light field, including spatial amplitude, phase and polarization, is quite challenging and indispensable for the generation of sophisticated optical lattices. Here, we propose and demonstrate a reconfigurable integrated full-dimensional optical lattice generator, i.e. a photonic emitting array (PEA) enabling reconfigurable and full-dimensional manipulation of optical lattices, in which 4×4 photonic emitting units (PEUs) with 64 thermo-optic microheaters are densely integrated on a silicon chip. By engineering each PEU precisely with independent and complete control of optical properties of amplitude, phase and polarization, various optical vortex lattices, cylindrical vector beam lattices, and vector vortex beam lattices can be generated and reconfigured in the far field. The demonstrated integrated optical lattice generator paves the way for the miniaturization, full-dimensional control and enhanced flexibility of complex light manipulation.**

The optical lattice is a fundamental physical phenomenon that can provide periodic potentials, finding a wealth of applications in neutral atoms cooling and trapping, quantum-state control, atomic clocks, microfluidic sorting, etc[1-4]. Apart from research in the field of atomic manipulation, the optical lattice has gained substantial interest in its specific kinds, classified by the fundamental properties of light waves such as spatial amplitude, phase, and polarization. For instance, partially coherent optical sources with periodic spatial coherence called optical coherence lattice, have intrinsic propagation characteristic that it can transfer the source coherence periodicity to the intensity profile periodicity in the far-field, showing application potentials in free-space optical communications, imaging and holography[5-11]. Furthermore, multiple optical vector beams with spatially inhomogeneous polarization distribution or optical vortex (OV) beams carrying orbital angular momentum (OAM) arranged in a network, known as cylindrical vector beam (CVB) lattice or OV lattice, is extensively studied[12, 13]. In comparison with single vector beam and OV, the CVB lattice and OV lattice are more sophisticated and flexible that can bring more attractive features. The increase in the number of CVBs in a CVB lattice means that it can realize fast particle sorting in optical trapping and simultaneously multiplex many channels in a space-division multiplexing (SDM) optical communication system[14-16]. In optical sensing, the deformation of OV lattice can be used for small linear displacement, rotation angle measurement, and even wavefront reconstruction[17, 18]. Thence, the generation of optical lattice is crucial in these advanced applications that has long been the focus of the attention.

To date, various optical lattice generation approaches have been demonstrated. One general technique utilizes the superposition of several element beams to create periodic light fields. For example, versatile OV lattices could be successfully produced by using multi-plane-wave interferometer or superimposed Laguerre–Gaussian (LG) beams with designed pairs of topological charge[19, 20]. Another technique uses spatial light modulators (SLMs) to shape the wavefront of light beams and then transform them into optical lattice, whose node status and structure are freely adjusted within a certain range[21-23]. Besides, there are some other extraordinary options to realize optical lattice generation such as Dammann gratings, holograms, and liquid crystals[24-30]. However, the universal methods to generate optical lattice based on bulk optics suffer from large volumes and long working distances that might limit their scalability and miniaturization in practical applications. Subsequently, the development of nanophotonic circuits based on versatile material platforms in last decades has given rise to the compact integration of optical lattice generators[31-33]. On silicon or InGaAsP/InP platforms, various optical lattice generators have been reported using specially designed nanostructures including microrings, tilted gratings, large-scale phase array, etc.[34-41]. Nevertheless, microrings resonators combined with gratings can be utilized to produce OV lattice, whose mode order and polarization state are limited by the structure of azimuthal gratings[34,35]. Tilted gratings can realize optical lattice generation with triangular grid, yet it does not achieve adjustability and reconfigurability[36]. Large-scale phased array is a potential solution to implement dynamic modulation of optical lattice, only the phase of light can be changed [37,38]. Thus, it still remains a great challenge to implement simultaneous and independent control of full dimensions of generated optical lattice in the previous state-of-the-art works.

In this work, we present a chip-scale reconfigurable integrated full-dimensional optical lattice generator with simultaneous, independent, and dynamic manipulation of spatial amplitude, phase and polarization of light fields based on the silicon platform. The proposed integrated optical lattice generator is a photonic emitting array (PEA) composed of 4×4 compact photonic emitting units (PEUs), where the amplitude, phase, and polarization of the near light field emitted from the unit can be modulated flexibly with the precise electrical tuning of the microheaters. By mixing of modulated near light fields generated from all the units, arbitrary complex radiation patterns with desired transverse spatial distributions in the optical lattice can be produced. In the experimental demonstration, various OV lattices, CVB lattices, and vector vortex beam (VVB) lattices are successfully generated in the far field using the designed and fabricated silicon photonic integrated circuit, showing favorable operation performance.

## Results

### Principles of integrated optical lattice generator

The concept and structure of the proposed integrated optical lattice generator based on the platform of silicon-on-insulator (SOI) is schematically illustrated in Fig. 1. In order to achieve various optical lattice generation containing periodic spatial amplitude, phase, and polarization distributions, full-dimensional manipulation of the light field is crucial. Figures 1a-1c depict the operation principle of phase, amplitude, and polarization modulation of light waves using integrated silicon photonic elements, respectively. As well known, on-chip optical phase can be controlled by the method of delay path or refractive index change. To realize an optical phase shift of $\varphi_p$, the thermo-optic microheater can be utilized to tune the refractive index of the silicon waveguide, and hence the optical phase is modulated, as shown in Fig. 1a. The general form of the phase-modulated light field $E_p$ can be expressed as

$$E_p = E_{in} \cdot e^{j\varphi_p} , \tag{1}$$

where $E_{in}$ is the electrical field of the input light. For optical amplitude modulation, as shown in Fig. 1b, the integrated Mach-Zehnder interferometer (MZI) is employed, where the amplitude of interfering light $A$ at the output side can be varied by tuning the phase difference $\varphi_a$ between two arms

$$A = \sqrt{\frac{1+\cos\varphi_a}{2}} . \tag{2}$$

As for the polarization modulation, any state on Poincaré sphere (PS) can be obtained by overlapping two orthogonal linear polarization states. Figure 1c shows a representative compact approach consisting of two orthogonally arranged waveguides, a two-dimensional (2D) vertical emission grating, and a thermo-optic microheater, which can implement on-chip polarization state control. Through changing the amplitude $A_x$, $A_y$ and phase difference $\varphi_{pol}$ between the fundamental TE modes in these two orthogonally arranged waveguides, the light emitted from the 2D grating has arbitrary polarization state, whose Jones vector $\mathbf{E_{pol}}$ can be written by

$$\mathbf{E_{pol}} = \begin{bmatrix} A_x \\ A_y e^{j\varphi_{pol}} \end{bmatrix}. \tag{3}$$

By combing these basic integrated silicon photonic elements, the concept of on-chip PEU achieving full-dimensional modulation of the light field is shown in Fig. 1d. The PEU consists of three major functional parts: a phase modulation to adjust overall phase $\varphi_p$ of the light arriving at the PEU, a $y$-junction splitter followed by two tunable MZIs to simultaneously modulate the amplitude $A_x$, $A_y$ of lights in two optical paths, and a 2D vertical emission grating with a heater to efficiently transfer the linearly polarized lights in two optical paths to synthetic optical beams in free space with polarization modulation. The Jones vector of the light emitted from a PEU can be expressed as follows

$$\mathbf{E_{PEU}} = F\left(\theta_x, \theta_y\right) \frac{e^{-jk_0|\mathbf{r}|}}{|\mathbf{r}|} \cdot E_p \cdot \mathbf{E_{Pol}} = F\left(\theta_x, \theta_y\right) \frac{e^{-jk_0|\mathbf{r}|}}{|\mathbf{r}|} \cdot E_{in} \cdot e^{j\varphi_p} \cdot \begin{bmatrix} A_x \\ A_y e^{j\varphi_{pol}} \end{bmatrix}, \tag{4}$$

where $\theta_x$ and $\theta_y$ are azimuth angles of pointing vector $\mathbf{r}$'s projections on the $xz$-plane and $yz$-plane, respectively. $F(\theta_x,\theta_y)$ is the radiation vector in the far field, which is related to the luminous features of the 2D grating. $k_0$ is the wave vector in free space. By assembling multiple PEUs to form an array (suppose $E_{in} = 1$), the schematic diagram of photonic emitted array (PEA) is shown in Fig. 1e, whose Jones vector is

$$\mathbf{E}_{\mathbf{PEA}} = F\left(\theta_x, \theta_y\right) \frac{e^{-jk_0|\mathbf{r}|}}{|\mathbf{r}|} \sum_{n_x=0}^{N-1} \sum_{n_y=0}^{N-1} e^{j\varphi_{p,n_x n_y}} \begin{bmatrix} A_{x,n_x n_y} \\ A_{y,n_x n_y} e^{j\varphi_{pol,n_x n_y}} \end{bmatrix} e^{j\Lambda k_0\left(n_x \sin\theta_x + n_y \sin\theta_y\right)}, \quad (5)$$

where *N* is the numbers of PEUs along the direction of *x*-axis or *y*-axis, and Λ is the period of adjacent PEUs (see supplementary information). Eq. (5) implies that the far-field of light radiated from PEA is the coherent addition of lights emitted from all the PEUs, where the number of radiation lobes is decided by period Λ, wave vector $k_0$, and azimuth angles $\theta_x$ and $\theta_y$. We assume PEUs arranged as a square ring in the PEA are utilized. Each PEU has the same amplitude and polarization modulation ( $A_{x,n_x n_y} = A_x$ , $A_{y,n_x n_y} = A_y$ , $\varphi_{pol,n_x n_y} = \varphi_{pol}$ ), while the phase difference between neighboring PEUs equals to $\pi l / 2\left(N-1\right)$ (*l* is an integer). When azimuth angles satisfy $\theta_x = p\lambda_0/\Lambda$ , $\theta_y = q\lambda_0/\Lambda$ (*p*, *q* = 0, ±1, ±2...), Eq. (5) can be changed to

$$\mathbf{E}_{\mathbf{PEA}}^{\mathbf{OV}}\left(p,q\right) = F\left(\frac{p\lambda_0}{\Lambda}, \frac{q\lambda_0}{\Lambda}\right) \frac{e^{-jk_0|\mathbf{r}|}}{|\mathbf{r}|} \cdot \frac{\sin\left(\frac{\pi l}{4}\right)}{\sin\left[\frac{\pi l}{4\left(N-1\right)}\right]} e^{j\Lambda k_0 (N-1)\Delta\theta \frac{1}{\sqrt{2}} \sin\left(\varphi + \frac{\pi}{4}\right)} \cdot \\ e^{j\frac{\pi l(4N-5)}{4(N-1)}} \cdot 4 j^{|l|} J_{|l|}\left(\frac{1}{\sqrt{2}} \Lambda k_0 \left(N-1\right)\Delta\theta\right) \begin{bmatrix} A_x \\ A_y e^{j\varphi_{pol}} \end{bmatrix} e^{jl\varphi}, \quad (6)$$

where $\Delta\theta$ is a slight offset to the azimuth angle, $\lambda_0$ is the wavelength in free space, $J_{|l|}$ represents the $|l|$-order Bessel functions of the first kind, and $\varphi$ is the azimuth angle in the *xy*-plane (see supplementary information). Eq. (6) predicts that the helical phase factors $e^{jl\varphi}$ simultaneously emerges at the positions of $\theta_x = p\lambda_0/\Lambda$ , $\theta_y = q\lambda_0/\Lambda$ , forming a rectangular OV lattice of order *l* with azimuth angle period of $\lambda_0/\Lambda$ . The density of OV lattice is determined by wavelength $\lambda_0$ and PEU period

Λ, where a larger PEU period can result in more OVs. This means that the distance between adjacent PEUs does not have to be designed to be small enough to restrain the side lobes, which is difficulty for photonic integrated circuits. Apart from OV lattice, more complex lights such as CVB lattice and VVB lattice can be produced by precisely varying the parameters of $\varphi_{p,n_x n_y}$ , $A_{x,n_x n_y}$ , $A_{y,n_x n_y}$ , $\varphi_{pol,n_x n_y}$ in all PEUs.

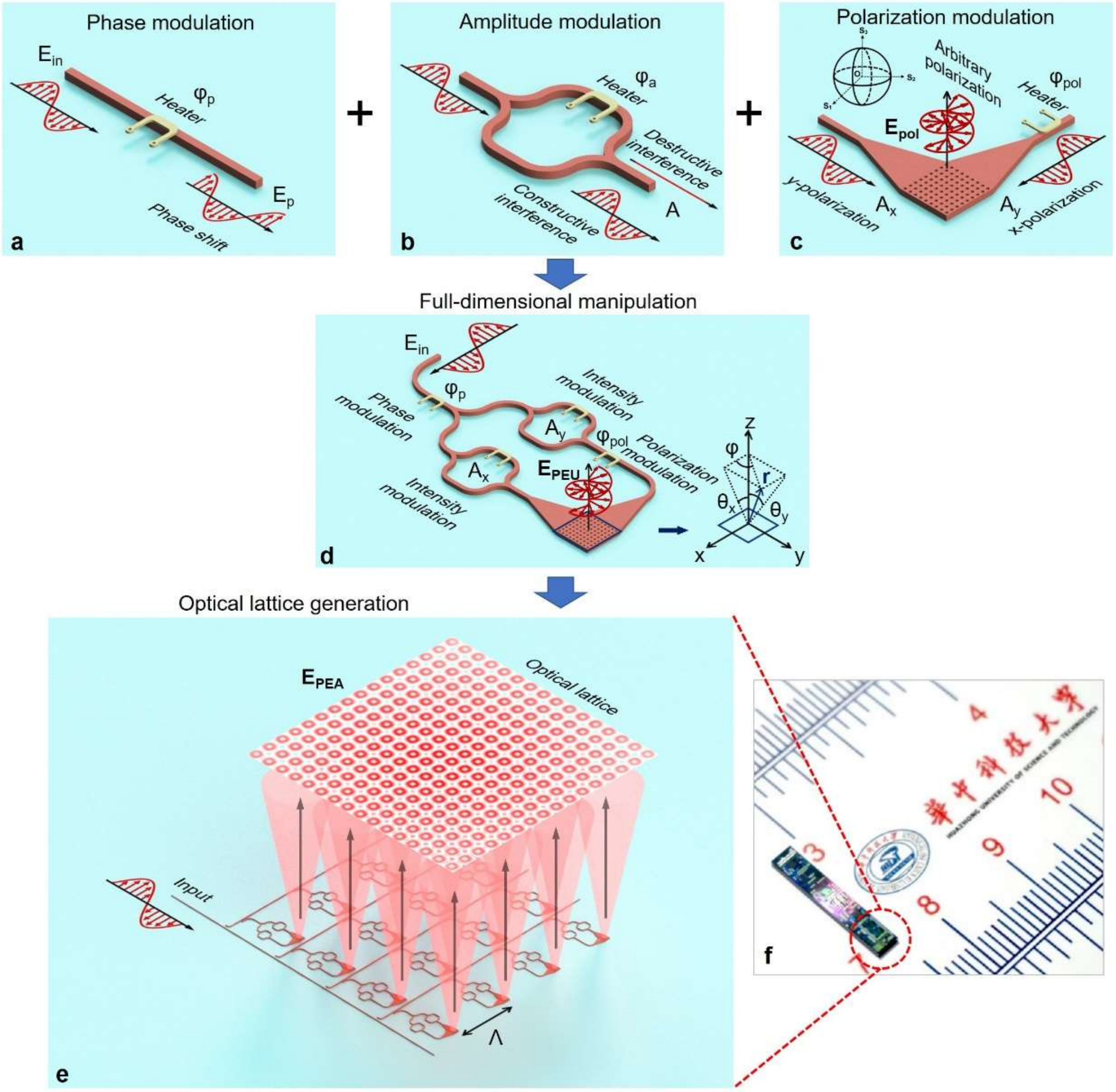

**Figure 1 | Schematic diagram of the proposed reconfigurable integrated full-dimensional optical lattice generator. a,** Phase, **b,** amplitude, and **c,** polarization modulation of light using integrated structures. **d,** Concept of full-dimensional manipulation of the light field using an integrated photonic emitting unit (PEU). **e,** Operation principle of optical lattice generation using integrated photonic emitting array (PEA) composed of 16×16 PEUs. **f,** Photograph of the fabricated reconfigurable integrated full-dimensional optical lattice generator.

Here, a preliminary verification has been carried out by designing and fabricating an integrated optical lattice generator based on a PEA with 4 × 4 PEUs and 64 thermo-optic microheaters. Figure 1f shows the photograph of the fabricated device, where the area of the PEA is less than 1 mm ×1 mm. Figures 2a-2c respectively show the fabricated device with wire bonding on printed circuit board (PCB), micrograph of top view of PEA, and the scanning electron microscope (SEM) image of a PEU. The optical lattice generator is fabricated on an SOI wafer with a 220-nm-thick top silicon layer and a 2-μm-thick buried oxide layer, where the etching depth of waveguides and 2D emitting gratings is 130 nm. The period between neighboring PEUs is 177 μm. The light in a single-mode fiber is input into the generator using a vertically coupled grating, then delivered equally to each PEU by 20 directional couplers with varying coupling length. Figure 2d displays the measured near-field intensity profile of the entire PEA, showing the nearly uniform emission across all of the 16 PEUs at the wavelength of 1550 nm. The measured amplitude and phase modulation characteristics of a PEU is systematically illustrated in Fig. 2e, where a polarizer is set above the 2D grating to filter out the linearly-polarized (LP) light coming from one MZI. One can see that the amplitude of light emitted from the 2D grating can be changed periodically with the increase of the applied voltages on the heater of MZI. While the phase response of the emitted light can cover 0-2π with the change of the applied voltages on the heater of phase modulation. In addition, the polarization modulation feature of a PEU is also verified in the experiment, as shown in Figs. 2f and 2g, where it requires three microheaters (two on MZIs and one on polarization modulation) to work together. To verify the polarization states of generated lights, polarizers and quarter wave plate are set above the 2D grating to measure the polarization components. Figure 2f plots the measured sinusoidal curves between normalized intensity and rotation angle of polarizer, indicating the successful generation of *x*-, *y*-, and 45°-polarized

LP light, respectively. Figure 2g plots the measured normalized intensity as functions of rotation angle of polarizer placed behind a quarter wave plate (QWP), showing the well generation of left-handed circularly polarized (LCP) light. The experimental results shown in Fig. 2 indicate that the PEU of the fabricated device has favorable performance of full-dimensional modulation of light.

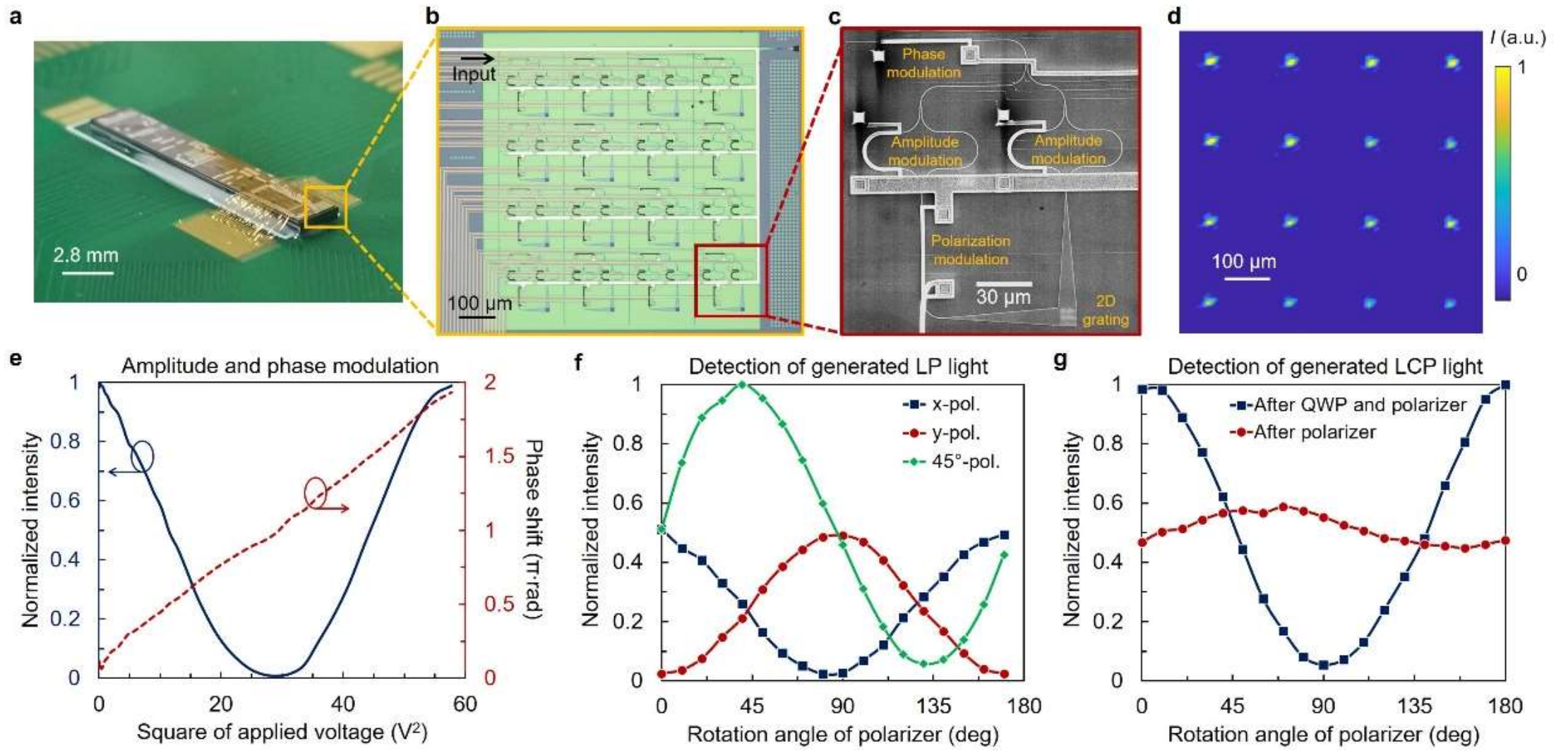

**Figure 2 | Fabricated optical lattice generator. a,** Fabricated device with wire bonding on PCB. **b,** Microscope image of top view of PEA. **c,** SEM image of a PEU. **d,** Measured near-field intensity profile of PEA. **e,** Measured amplitude and phase modulation responses of a PEU by tuning the microheaters. **f** and **g,** LP and LCP light generation in a PEU.

## Amplitude and phase manipulation

The amplitude and phase modulation properties of entire PEA of the fabricated optical lattice generator are investigated, as shown in Fig. 3. First, we simulate and demonstrate a direct measurement of the far-field intensity patterns of three neighboring PEUs, whose emitted lights have the same amplitude, phase, and polarization modulation. As depicted in Figs. 3a-3d, different near-field luminous positions in three neighboring PEUs will lead to distinctive radiation patterns at the wavelength of 1550 nm. For instance, two horizontally adjacent PEUs can produce a horizontally spaced pattern, as shown in Fig. 3a. While a similar vertically interference pattern is produced with two vertically adjacent PEUs, as shown in Fig. 3b. Furtherly, the larger PEU spacing with the diagonal distribution results in more dense interference fringes, as shown in Fig. 3c. When three PEUs are emitting synchronously, periodical fringes become an array of bright spots, as shown in Fig.

3d. Moreover, we investigate the far-field intensity patterns of four neighboring PEUs with different phase modulations. To verify the phase distribution of generated lights, another Gaussian beam is added on the far-field intensity profiles to obtain the tilt interferograms, which are compared with simulated phase patterns. One can see that the far-field intensity pattern of adjacent four PEUs with the same phase modulation has a rectangular network of bright spots, whose phase distribution is similar to Gaussian beam, as shown in Fig. 3e. While the far-field tilt interferogram of adjacent four PEUs with gradient varying phase modulation (from 0 to $2\pi$) has multiple fork patterns, denoting the generation of an array of phase singularities, as shown in Fig. 3f. The measured results in Fig. 3 match the simulation ones well, meaning that the PEA can generate light beam with periodic structures by utilizing amplitude and phase modulation, which is crucial to produce more sophisticated optical lattice.

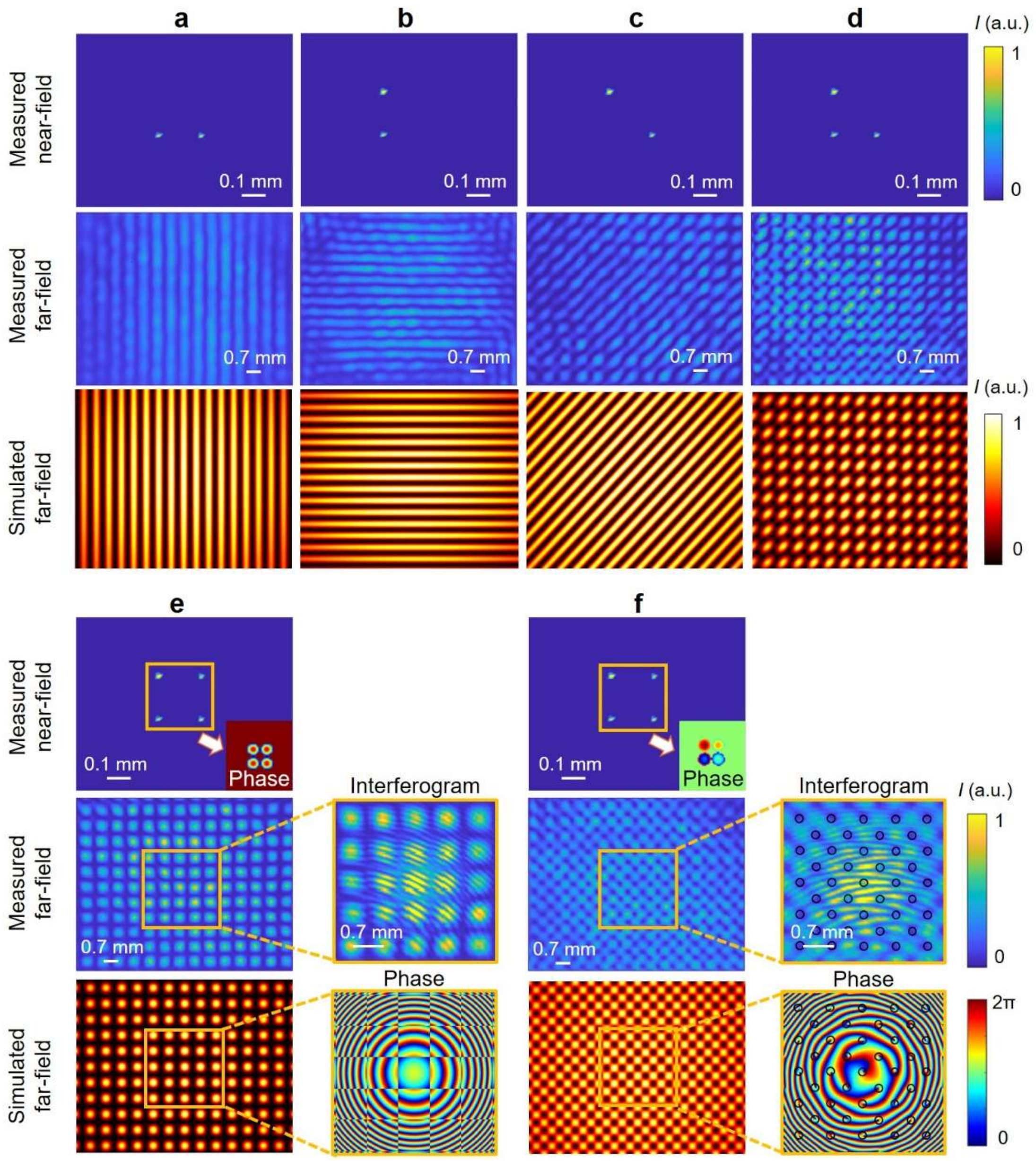

**Figure 3 | Amplitude and phase manipulation of the integrated optical lattice generator. a-d,** Simulated and experimentally measured intensity profiles produced by the generator: (**a-c**) the interferences of two adjacent PEUs and (**d**) three adjacent PEUs. Comparison of the intensity profiles of four PEUs with (**e**) the same phase modulation and (**f**) with gradient varying phase distribution.

## OV lattice generation

Afterwards, we utilize the amplitude and phase modulation of PEA to generate OV lattice. To obtain OV with high purity, the initial phase distribution of each PEU should be designed and controlled very precisely. Figure 4a shows the simulated OV lattice generated by the PEA. Here, 8 and 12 PEUs in a square with phase tuning are chosen to generate *x*-polarized OV lattice with $OAM_{-1}$ and $OAM_{+2}$, respectively.

As shown in Fig. 4a, the first column displays the phase modulation of chosen PEUs with azimuthal clockwise change from 0 to 2π ($OAM_{-1}$) and with azimuthal counterclockwise change from 0 to 4π ($OAM_{+2}$). The second and third columns respectively show far-field intensity profiles of generated OV lattice and their zoom-in views, from which one can see an array of doughnut shape patterns with dark spot in the center. The fourth column plots the phase distribution of far-field light with an array of phase singularities, verifying the generation of OV lattice. Figure 4b shows the measured OV lattice generated by the fabricated PEA. Here, 8 PEUs in a square with azimuthal clockwise and counterclockwise phase change from 0 to 2π are chosen to generate *x*-polarized OV lattice with $OAM_{-1}$ and *y*-polarized OV lattice with $OAM_{+1}$, respectively. 12 PEUs in a square with azimuthal counterclockwise phase change from 0 to 2π are chosen to generate *y*-polarized OV lattice with $OAM_{+2}$. The first column shows the measured near-field intensity profiles of PEA. The second and third columns respectively show measured far-field intensity profiles of generated optical patterns and their zoom-in views, which are composed of an array of doughnut shape patterns. To verify the phase distribution of generated optical patterns, another Gaussian beam is coaxially added on them to produce interferograms, as shown in the fourth column. One can indicate that the measured interferograms have helical interference fringes, confirming the generation of OV lattice. To evaluate the quality of the generated optical lattice, we measure the phase purity of OV in the lattice by calculating the overlapping integrals between the reconstructed phase profile and the theoretical one. The reconstructed phase profile of OV can be obtained by its tilt interferogram. The measured phase purity of generated OV of *x*-polarized $OAM_{-1}$ versus optical wavelength (from 1532.5 to 1595.0 nm) is depicted in Fig. 4c, whose average value is about 0.82. The results denote that the designed PEA can generate broadband linearly polarized OV lattice with relatively high phase purity.

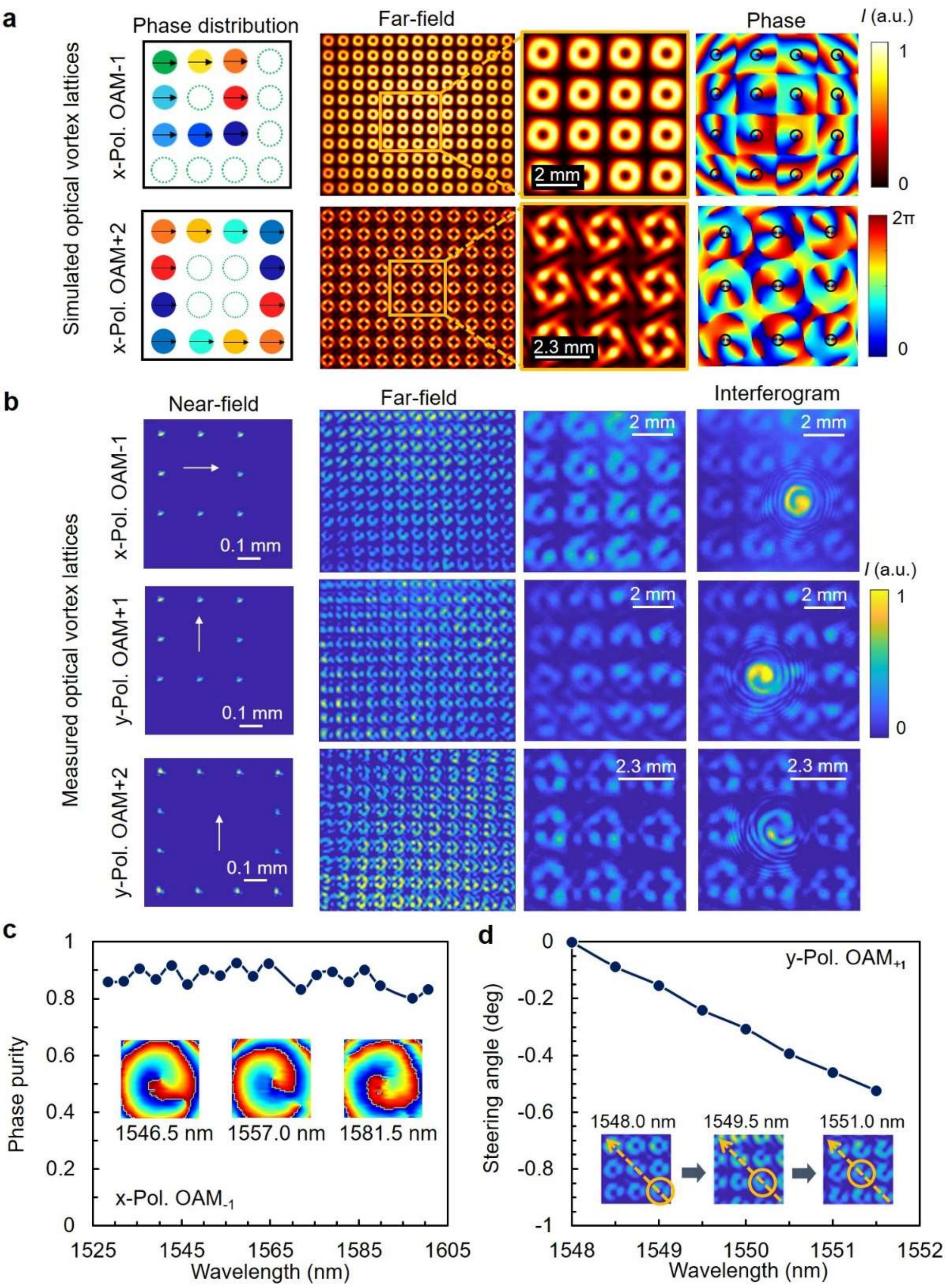

**Figure 4 | Generation of OV lattices. a,** Simulated intensity and phase profiles of generated OV lattices with $OAM_{-1}$ and $OAM_{+2}$, respectively. **b,** Experimentally measured intensity profiles and interferograms of OV lattices with *x*-polarized $OAM_{-1}$, *y*-polarized $OAM_{+1}$ and *y*-polarized $OAM_{+2}$, respectively. **c,** Calculated mode purity of the measured OV with *x*-polarized $OAM_{-1}$ as a function of the optical wavelength. Insets are the reconstructed phase profiles. **d,** Experimentally measured *x*/*y*-axis steering angle of OV lattice with *y*-polarized $OAM_{+1}$ as a function of the optical wavelength. Insets are the far-field intensity distributions of OV lattice under different optical wavelengths.

Likewise, we study the steering features of the OV lattice generated by the fabricated PEA. As well known, the optical beam steering direction depends on the phase difference of the PEUs. One can see from Fig. 1e that there exists a fixed optical path difference relative to the period Λ between the adjacent PEUs in both *x*- and *y*-directions. When optical wavelength is varying, the phase difference between adjacent PEUs induced by the optical path difference is also changing, leading to the steering of generated optical beams in the far field. Figure 4d shows the measured *x*/*y*-axis steering angle of *y*-polarized OV lattice with $OAM_{+1}$ as function of optical wavelength, where the insets display the intensity of OV lattice with different wavelength. One can indicate that OV lattice moves diagonally, when the wavelength varies from 1548 to 1551.5 nm. Since the OVs in the generated lattice are spaced about 0.52° in both *x*- and *y*- directions, the maximum wavelength tuning range is about 3.5 nm. It means that one cannot tell the intensity patterns of OV lattice if the wavelength changes beyond 3.5 nm. The radiation lobes will overlap when the wavelength varies from 1548 to 1551.5 nm.

**CVB and VVB lattice generation**

By combining the amplitude, phase and polarization modulation of each PEU in the fabricated generator, we experimentally demonstrate the CVB lattice generation. Generally, the Jones vector of CVBs with circularly symmetric LP distribution can be expressed as

$$\mathbf{E}_{\mathbf{CVB}}^{\mathbf{LP}} = \begin{bmatrix} \cos\left(\varphi_0 + l_p\varphi\right) \\ \sin\left(\varphi_0 + l_p\varphi\right) \end{bmatrix} = \frac{1}{\sqrt{2}}\left(e^{j\varphi_0}e^{jl_p\varphi}\mathbf{R} + e^{-j\varphi_0}e^{-jl_p\varphi}\mathbf{L}\right), \quad (7)$$

where **L** and **R** are LCP and right-handed circularly polarized (RCP) components, respectively, whose Jones vectors are $\left[1 \quad \pm j\right]^T / \sqrt{2}$ . $\varphi_0$ is a fixed angle determining the polarization orientation of the CVBs. $l_p$ is the order of the CVBs. To generate the desired CVB lattice, the polarization distribution of emitted lights of all PEUs should be adjusted to match Eq. (7), while the amplitude and phase values of them maintain the same. Figure 5a shows the generated CVB lattice using the designed PEA with $l_p$ = +1 and $\varphi_0$ = 0° (radially polarized CVB). Here 8 PEUs in a square are used to emit lights with radially polarized distribution. The first row displays the measured near-field intensity profiles of PEA, where a polarizer with varying orientation angle is applied to verify the polarization state of emitted lights. After the lights pass through the polarizer with orientation angle of 0° and 90°, bright spots on the vertical and horizontal line in the middle are greatly attenuated, respectively. The similar phenomenon emerges when the orientation angle of polarizer is respectively switched to 45° and 135°, proving that the polarization distribution of PEA matches the radially polarized CVB well. The second and third rows respectively show the simulated and measured far-field intensity profiles of PEA, whose polarization features are also analyzed by a polarizer. The orientation angle of measured dark fringes is vertical to that of the polarizer, fitting well with the simulation results. To furtherly confirm the polarization state of generated optical lattice, circularly polarized Gaussian beams are applied to coaxially interfere the far-field lights since a CVB with order $l_p$ can be regarded as a superposition of an OV with RCP $OAM_{+l_p}$ and an OV with LCP $OAM_{-l_p}$, as illustrated in Eq. (7). When the generated optical lattice is coaxially interfered with an LCP (RCP) Gaussian beam followed by a circularly polarized filter composed of a QWP and a polarizer, helical interference fringes appear in the interferogram, verifying the components of LCP $OAM_{-1}$ (RCP $OAM_{+1}$), as shown in Fig. 5a. Moreover, generation of CVB lattice with other states is also investigated in the experiment. Figure 5b shows a Poincaré sphere with $l_p$ = +1, where the poles represent two OV beams with uniform amplitude but opposite helicity and polarization (RCP $OAM_{+1}$ at the north pole and LCP $OAM_{-1}$ at the south pole). The CVBs with four different states are presented on the equator of the sphere, whose polarization orientation angle $\varphi_0$ are 0°, 45°, 90°, and 135°, respectively. Utilizing 8 PEUs in a square with specific polarization modulation, CVB lattice with $\varphi_0$ = 45°, 90°, and 135° are respectively generated, as depicted in Figs. 5c-e. After passing through the polarizer, the doughnut-shaped intensity profiles of CVBs in the lattice become two-lobed, where the orientation angle of dark fringes is 45°, 0°, and -45° relative to that of polarizer, respectively, verifying the polarization states of generated CVBs.

Apart from CVB lattice, more sophisticated optical lattice carrying specific phase and polarization distribution called VVB lattice generation is also demonstrated using the fabricated PEA. Similar to the expression in Eq. (7), the Jones vector of VVBs with circularly symmetric LP distribution can be written by

$$\mathbf{E}_{\mathbf{VVB}}^{\mathbf{LP}} = e^{jl_o\varphi}\begin{bmatrix}\cos\left(\varphi_0 + l_p\varphi\right) \\ \sin\left(\varphi_0 + l_p\varphi\right)\end{bmatrix} = \frac{1}{\sqrt{2}}\left(e^{j\varphi_0}e^{j\left(l_o+l_p\right)\varphi}\mathbf{R} + e^{-j\varphi_0}e^{j\left(l_o-l_p\right)\varphi}\mathbf{L}\right), \quad (8)$$

where $l_p$ and $l_o$ are the orders of polarization state and OV of the VVBs, respectively. $\varphi_0$ is the fixed orientation angle of the polarization state of the VVBs. Figure 5f shows the generated VVB lattice with $l_p$ = +1, $l_o$ = +1, and $\varphi_0$ = 0°. Here 12 PEUs in a square is used to emit lights with helical phase and radially polarized distribution. The first and second rows respectively display the simulated and measured far-field intensity of generated VVB lattice, whose linearly and circularly polarized components are analyzed by a polarizer and a QWP. Adding a polarizer with orientation angle of 0°, 45°, 90° and 135° on the path of the light, respectively, new dark fringes appear in the intensity profiles of VVB lattice, whose orientation angle is perpendicular to that of the polarizer. Besides, a VVB can be seen as a combination of RCP $OAM_{lo+lp}$ and LCP $OAM_{lo-lp}$, as described in Eq. (8). A circularly polarized filter (a QWP followed by a polarizer) is used to verify the circularly polarized properties. When the circularly polarized filter is set to LCP-passing, there emerges a bright-spot lattice. While the circularly polarized filter becomes RCP-passing, the VVB lattice is changed to doughnut-shaped intensity array, whose tilt interferogram contains fork structures with 2 tines. The results indicate that the light is composed of an array of RCP $OAM_{+2}$ and LCP $OAM_0$, verifying the generation of VVB lattice with $l_p$ = +1, $l_o$ = +1, and $\varphi_0$ = 0°.

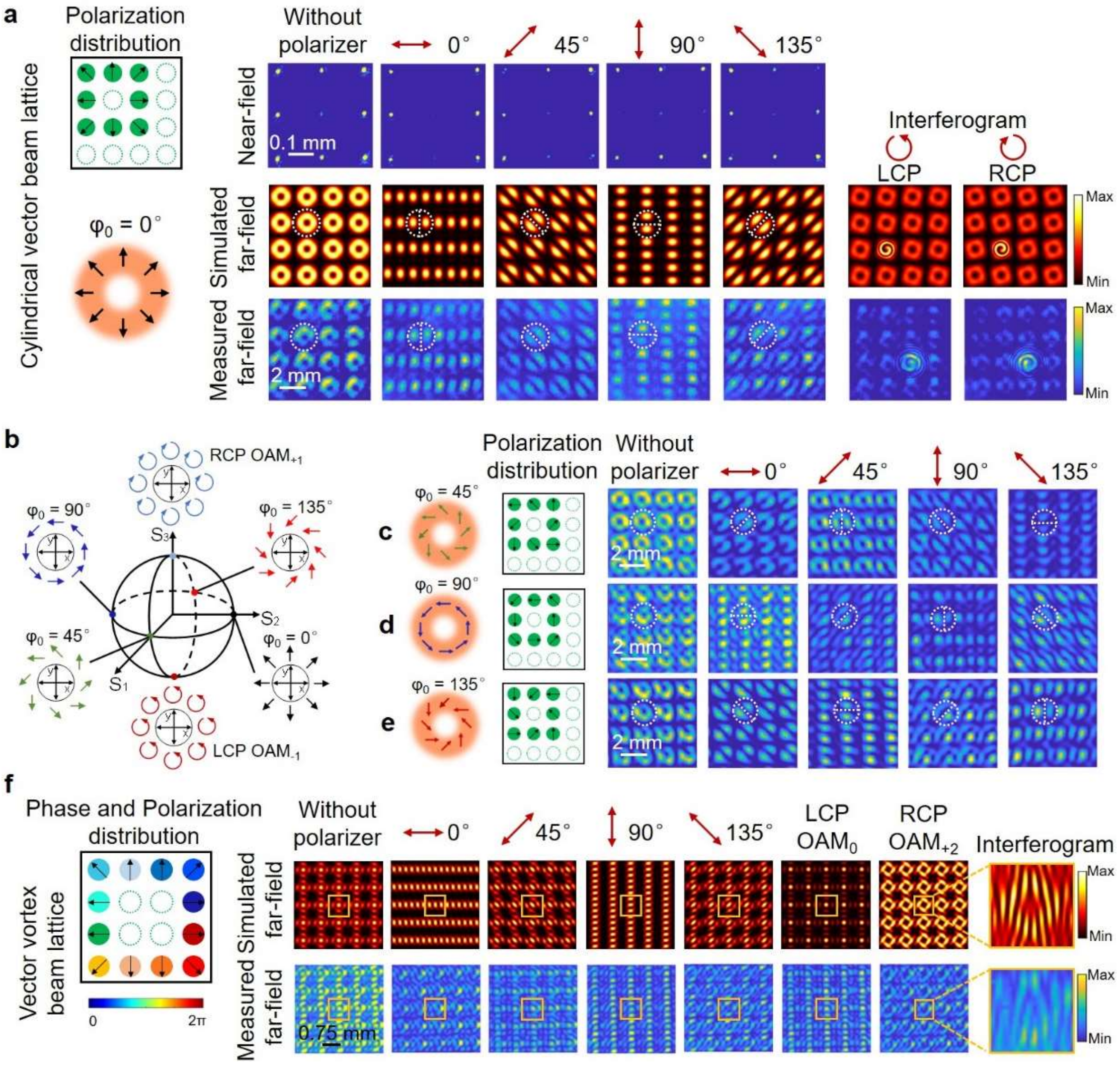

**Figure 5 | Generation of CVB lattices and VVB lattices. a,** Generation of CVB lattice with $l_p$ = +1 and $\varphi_0$ = 0° using the PEA. The first row of inserts is experimentally measured near-field intensity profiles of 8 PEUs. The second and third rows of inserts are simulated and experimentally measured far-field intensity profiles of PEA. **b,** A Poincaré sphere with $l_p$ = +1, on which the states of CVB are investigated in the experiment. **c-e,** Generation of CVB lattices with $l_p$ = +1 and $\varphi_0$ = 45°, 90°, and 135° using the PEA, respectively. Inserts are experimentally measured far-field intensity profiles of PEA. **f,** Generation of VVB lattices with $l_p$ = +1, $l_o$ = +1 and $\varphi_0$ = 0° using the PEA. The inserts are simulated and experimentally measured far-field intensity profiles of PEA.

## Discussion

Our integrated optical lattice generator, composed of an array of reconfigurable PEUs, can provide amplitude, phase and polarization modulation of light beams in the far field. This extends the flexibility of light manipulation greatly and achieves the real full-dimensional molding of optical lattice on silicon platform. The proposed device also exploits the larger PEU period to generate denser optical lattice, which greatly reduces the design difficulty existed in current nanophotonic arrays. We have experimentally demonstrated the controlled generation of various optical lattices, including OV lattices with high purity, as well as CVB lattices and VVB lattice. By varying optical wavelength, the functionality of OV lattice steering is also realized, providing more manipulation method. The multi-dimensional tunability and the arrayed PEUs enable the PEA to generate arbitrary sophisticated optical lattice in the far field dynamically. Further, this photonic integrated circuit may combine with semiconductor lasers, on-chip high-speed modulators, and integrated amplifiers to package together to meet demand in optical manipulation, sensing, imaging, and communication[42-44].

## Methods

**Photonic emitting unit design.** A common SOI wafer with 220 nm top silicon layer and buried oxide layer is used in our design. In each PEU, multimode interference (MMI) based Y-splitters are used to split and combine lights in the MZIs. A 2D grating is chose to combine and emit lights of two orthogonal LP states. Supplementary Fig. S1 respectively shows the microscope image of a PEU, SEM image of a 2D grating, and a Y-splitter. The 2D grating with an etching pattern of cylinder array is explored with etching depth of 130 nm, cylinder radius of 370 nm, and pitch of 570 nm. The size of 2D grating is chosen to be 10 μm ×10 μm.

**Device fabrication.** The designed chip is fabricated on an SOI wafer with 220-nm-thick top silicon layer and a 2-μm-thick buried oxide layer ($SiO_2$) (see supplementary Fig. S2). In the fabrication, the layout pattern is defined using 248 nm deep ultraviolet (DUV) lithography and inductively coupled plasma (ICP) etching. Waveguides outlines are etched fully down 220 nm to the buried oxide layer, while the vertical grating couplers are shallow etched down nominally 70 nm and the 2D gratings are shallow etched 130 nm. After that, a 2-μm-thick $SiO_2$ layer is deposited on the top by plasma enhanced chemical vapor deposition (PECVD), covering the whole device as the upper cladding. Then, 1.2-μm-thick metal Titanium nitride (TiN) with a length of 50 μm is deposited on the top of the waveguides, serving as the heaters for thermo-optic tuning. As shown in Fig. S2, the fabrication was done at a foundry MPW run, following the standard CMOS-compatible fabrication process.

**Experimental configuration.** In the experiment, a tunable continuous wave laser source is used to scan the fabricated device working at different wavelengths. The experimental setup for characterizing the fabricated chip is shown in supplementary Fig. S3. The laser output is divided into two branches via a 3-dB coupler. One branch with its polarization controlled by a polarization controller (PC) is used to excite the fundamental $TE_0$ mode guided in the fabricated chip via a fan-grating (vertical coupling). With the proper control of voltage source array (VSA), various optical lattices can be generated. The light fields emitted from the PEA on the chip firstly propagate vertically, and then transmit horizontally after a mirror. The near-field and far-field intensity distribution measurements can be realized by employing a three-lens optical system, where the focus lengths of Lens1, Lens2, and Lens3 are 50 mm, 150 mm and 200 mm, respectively. Followed by a pinhole, a QWP and a polarizer, optical lattices with anisotropic polarization distributions can be characterized. The other branch serves as a referenced Gaussian beam. The power and polarization state of the reference Gaussian beam are adjusted by utilizing a PC, a variable optical attenuator (VOA), and a collimator. The generated optical lattice and the referenced Gaussian beam with the same polarization, similar power and beam size are combined through a beam splitter (BS) to produce the interferogram. The wavelength of the laser can be tuned from 1500 to 1630 nm. The attenuation tuning range of the VOA is from 2 to 60 dB. A camera is applied to monitor the intensity profiles of generated optical lattices and their corresponding interferograms and polarization distributions.

**Acknowledgments**

This work was supported by the National Key R&D Program of China (2019YFB2203604), the National Natural Science Foundation of China (NSFC) (62125503), the Key R&D Program of Guangdong Province (2018B030325002), the Key R&D Program of Hubei Province of China (2020BAB001, 2021BAA024), and the Science and Technology Innovation Commission of Shenzhen (JCYJ20200109114018750).

**Author contributions**

J.W. developed the concept and conceived the idea of the work. J.D. and S.Z. performed the theoretical analyses. J.D. and J.W. designed the photonic integrated circuits. S.Z. and J.D. fabricated the photonic integrated circuits, performed the experiments, data analyses, and simulations. X.C., J.Z., Z.W., and Y.L. provided experimental supports. S.Z., J.D. and J.W. wrote the manuscript with inputs from all co-authors. J.W. finalized the paper. J.W. supervised the project.

**Additional information**

Competing financial interests: The authors declare no competing financial interests.

**Data availability**

All the findings of this study are available in the main text or the Supplementary Information. The raw data are available from the corresponding author upon reasonable request.

## References

1. Ushijima, I., et al., Cryogenic optical lattice clocks. *Nature Photonics* **9**, 185 (2015).
2. Tackmann, G., et al., Laser Controlled Tunneling in a Vertical Optical Lattice. *Physical Review Letters* **106**, 213002(2011).
3. MacDonald, M.P., G.C. Spalding and K. Dholakia, Microfluidic sorting in an optical lattice. *Nature* **426**, 421(2003).

4. Jessen, P.S. and I.H. Deutsch, Quantum-state control in optical lattices. *Physical Review A* **57**, 1972(1998).

5. Ricklin, J.C. and F.M. Davidson, Atmospheric turbulence effects on a partially coherent Gaussian beam: implications for free-space laser communication. *Journal of the Optical Society of America A* **19**, 1794(2002).

6. Takeda, M., et al., Coherence holography. *Optics Express* **13**, 9629(2005).

7. Chen, Y., S.A. Ponomarenko and Y. Cai, Experimental generation of optical coherence lattices. *Applied Physics Letters* **109**, 061107(2016).

8. Clark, J.N., et al., High-resolution three-dimensional partially coherent diffraction imaging. *Nature Communications* **3**, 993(2012).

9. Ma, L. and S.A. Ponomarenko, Optical coherence gratings and lattices. *Optics Letters* **39**, 6656(2014).

10. Auñón, J.M. and M. Nieto-Vesperinas, Partially coherent fluctuating sources that produce the same optical force as a laser beam. *Optics Letters* **38**, 2869(2013).

11. Liang, C., et al., Perfect optical coherence lattices. *Applied Physics Letters* **119**, 131109(2021).

12. Guo, L., et al., Generation of vector beams array with a single spatial light modulator. *Optics Communications* **490**, 126915(2021).

13. Eastwood, S.A., et al., Phase measurement using an optical vortex lattice produced with a three-beam interferometer. *Optics Express* **20**, 13947(2012).

14. Bhebhe, N., et al., A vector holographic optical trap. *Scientific Reports* **8**, 17387(2018).

15. Zhao, Y. and J. Wang, High-base vector beam encoding/decoding for visible-light communications. *Optics Letters* **40**, 4843(2015).

16. Fang, J., et al., Spin-Dependent Optical Geometric Transformation for Cylindrical Vector Beam Multiplexing Communication. *ACS Photonics* **5**, 3478(2018).

17. Masajada, J., Small-angle rotations measurement using optical vortex interferometer. *Optics Communications* **239**, 373(2004).

18. Masajada, J., et al., Vortex points localization problem in optical vortices interferometry. *Optics Communications* **234**, 23(2004).

19. Franke-Arnold, S., et al., Optical ferris wheel for ultracold atoms. *Optics Express* **15**, 8619(2004).

20. O Holleran, K., M.J. Padgett and M.R. Dennis, Topology of optical vortex lines formed by the interference of three, four, and five plane waves. *Optics Express*, **14**, 3039(2006).

21. Li, X., et al., Close-packed optical vortex lattices with controllable structures. *Optics Express* **26**, 22965(2018).

22. Li, L., et al., Generation of optical vortex array along arbitrary curvilinear arrangement. *Optics Express* **26**, 9798(2018).

23. Wang, J., et al., Terabit free-space data transmission employing orbital angular momentum multiplexing. *Nature Photonics* **6**, 488(2012).

24. Yu, J., et al., Generation of dipole vortex array using spiral Dammann zone plates. *Applied Optics* **51**, 6799(2012).

25. Bortolozzo, U., et al., Harnessing Optical Vortex Lattices in Nematic Liquid Crystals. *Physical Review Letters* **111**, 093902(2013).

26. Ladavac, K. and D.G. Grier, Microoptomechanical pumps assembled and driven by holographic optical vortex arrays. *Optics Express* **12**, 1144(2004).

27. Son, B., et al., Optical vortex arrays from smectic liquid crystals. *Optics Express* **22**, 4699(2014).

28. Piccardo, M., et al., Vortex laser arrays with topological charge control and self-healing of defects. *Nature Photonics* **16**, 359(2022).

29. Liu, S., et al., 3D Engineering of Orbital Angular Momentum Beams via Liquid-Crystal Geometric Phase. *Laser & Photonics Reviews* **16**, 2200118(2022).

30. Jin, J., et al., Multi-Channel Vortex Beam Generation by Simultaneous Amplitude and Phase Modulation with Two-Dimensional Metamaterial. *Advanced Materials Technologies* **2**, 1600201(2017).

31. Bogaerts, W., et al., Basic structures for photonic integrated circuits in

Silicon-on-insulator. *Optics Express* **12**, 1583(2004).

32. Coldren, L.A., S.W. Corzine and M.L. Mashanovitch, Diode lasers and photonic integrated circuits. 2012: *John Wiley & Sons*.

33. R., N., et al., InP Photonic Integrated Circuits. *IEEE Journal of Selected Topics in Quantum Electronics* **16**, 1113(2010).

34. Zhang, J., et al., An InP-based vortex beam emitter with monolithically integrated laser. *Nature Communications* **9**, 2652(2018).

35. Cai, X., et al., Integrated Compact Optical Vortex Beam Emitters. *Science* **338**, 363(2012).

36. Du, J. and J. Wang, Chip-scale optical vortex lattice generator on a silicon platform. *Optics Letters* **42**, 5054(2017).

37. Sun, J., et al., Large-scale nanophotonic phased array. *Nature* **493**, 195(2013).

38. Poulton, C.V., et al., Large-scale silicon nitride nanophotonic phased arrays at infrared and visible wavelengths. *Optics Letters* **42**, 21(2017).

39. Miao, P., et al., Orbital angular momentum microlaser. *Science* **353**, 464(2016).

40. Li, H., et al., Orbital angular momentum vertical-cavity surface-emitting lasers. *Optica* **2**, 547(2015).

41. Qiao, X., et al., Higher-dimensional supersymmetric microlaser arrays. *Science* **372**, 403(2021).

42. Liu, Y., et al., A photonic integrated circuit-based erbium-doped amplifier. *Science* **376**, 1309(2022).

43. Powell, K., et al., Integrated silicon carbide electro-optic modulator. *Nature Communications* **13**, 1851(2022).

44. Xiang, C., et al., Laser soliton microcombs heterogeneously integrated on silicon. *Science* **373**, 99(2021).

# Supplementary Information for

# Reconfigurable integrated full-dimensional optical lattice generator

Shuang Zheng[1,2]†, Jing Du[1,2]†, Xiaoping Cao[1,2], Jinrun Zhang[1,2], Zhenyu Wan[1,2], Yize Liang[1,2], Jian Wang[1,2]*

[1] Wuhan National Laboratory for Optoelectronics and School of Optical and Electronic Information, Huazhong University of Science and Technology, Wuhan 430074, Hubei, China.

[2] Optics Valley Laboratory, Hubei, Wuhan 430074, China

† These authors contributed equally to this work.

*Corresponding author: jwang@hust.edu.cn

## 1. Photonic emitting unit (PEU) design

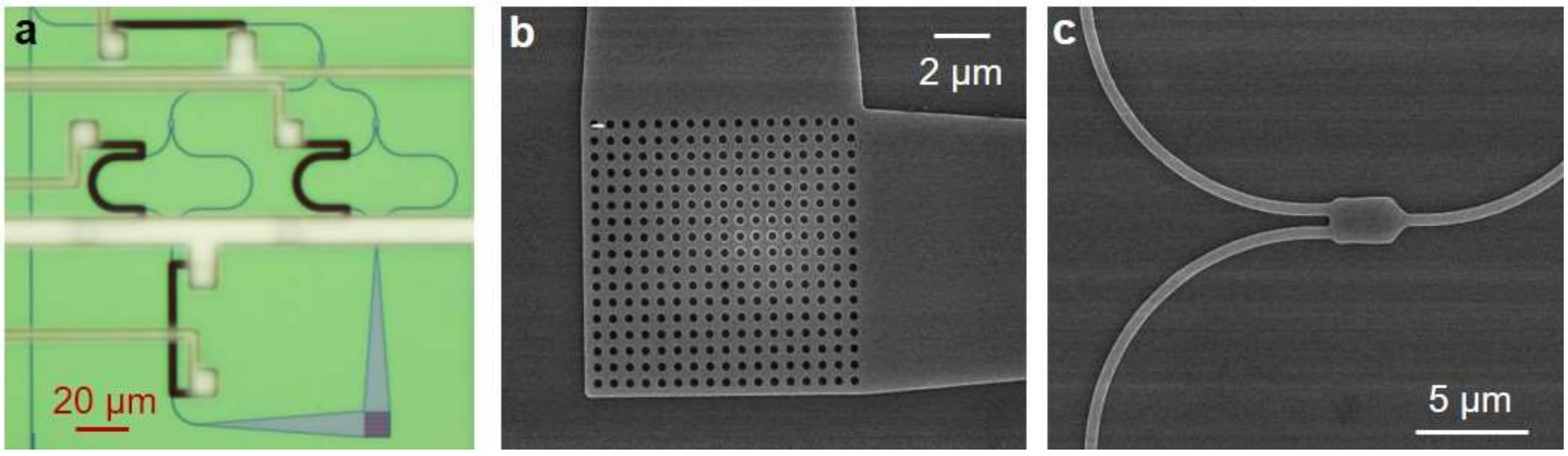

**Figure S1 | PEU design. a,** Microscope image of a PEU. **b,c,** Measured SEM image of (**b**) a 2D grating and (**c**) a Y-splitter.

## 2. Device fabrication process

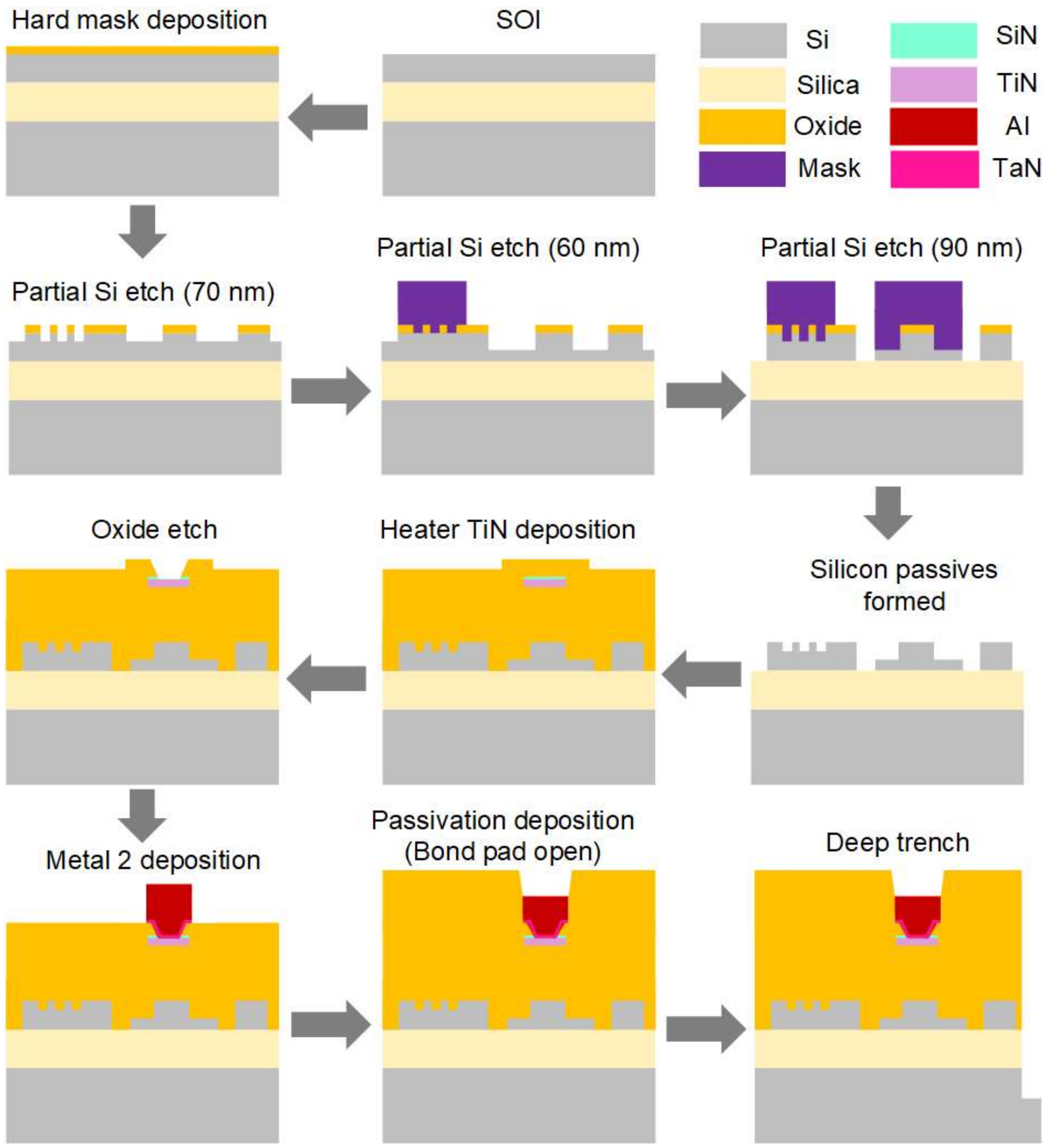

**Figure S2 | Fabrication process of integrated optical lattice generator.**

## 3. Experimental configuration

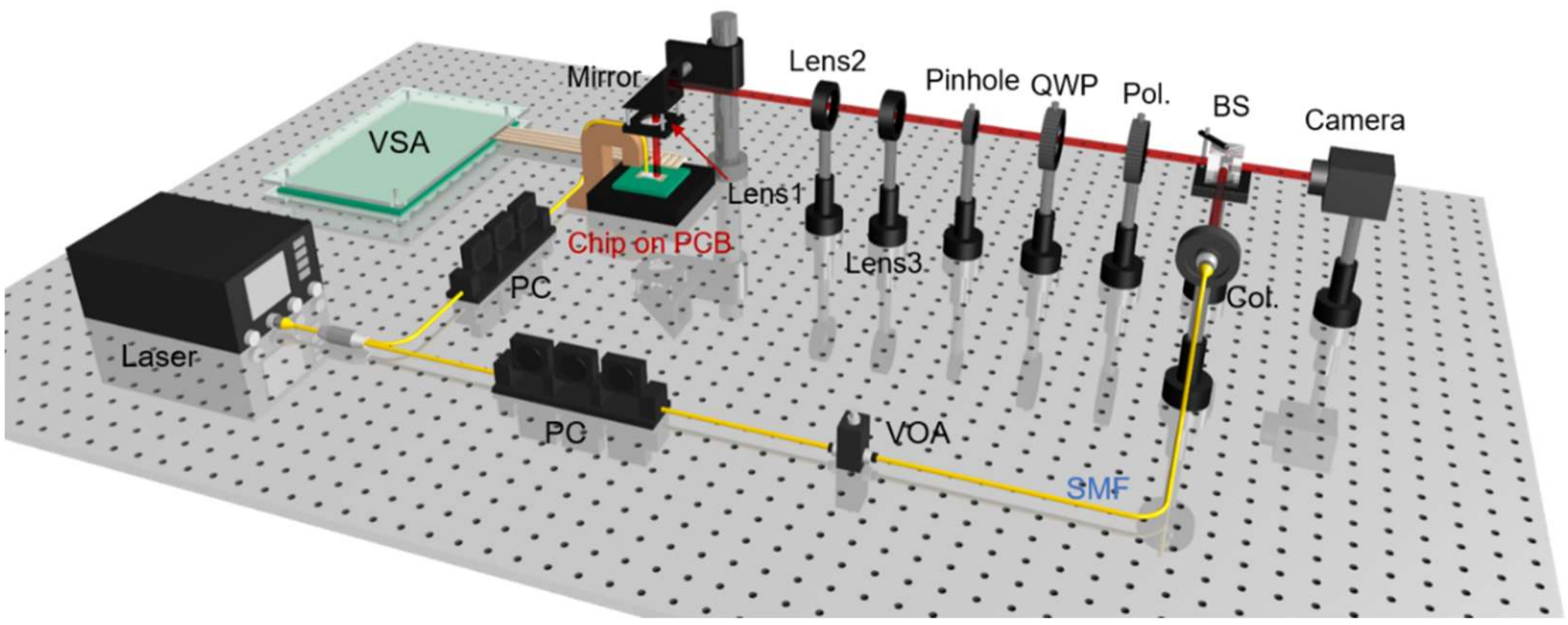

**Figure S3 | Experimental setup.** VSA, voltage source array; PC, polarization controller; VOA, variable optical attenuator; Col., collimator; Pol., polarizer; QWP, quarter-wave plate; BS, beam splitter; SMF: single-mode fiber.

## 4. Theory for the optical lattice generation

The theory model of PEA is illustrated in Fig. S4. Figure S4a shows the light diffraction model of an ideal rectangular array composed of $N_x \times N_y$ PEUs. The Jones vector of the far-field light emitted from a PEA can be expressed as follows[1]

$$\begin{bmatrix} E_{x,PEA}(\mathbf{r}) \\ E_{y,PEA}(\mathbf{r}) \end{bmatrix} = \sum_{n_x=0}^{N_x-1} \sum_{n_y=0}^{N_y-1} F_{n_x n_y}(\theta_x, \theta_y) e^{j\varphi_{p,n_x n_y}} \begin{bmatrix} A_{x,n_x n_y} \\ A_{y,n_x n_y} e^{j\varphi_{pol,n_x n_y}} \end{bmatrix} \frac{e^{-jk_0 |\mathbf{r} - \mathbf{u}_{n_x n_y}|}}{|\mathbf{r} - \mathbf{u}_{n_x n_y}|}, \quad \text{(S1)}$$

where $E_{x,PEA}(\mathbf{r})$ and $E_{y,PEA}(\mathbf{r})$ are scalar electric fields of *x*-polarization and *y*-polarization at the point **r** in the far-field, respectively. $F_{n_x n_y}(\theta_x, \theta_y)$ is the radiation vector of $PEU_{n_x n_y}$ in the far-field, in which $\theta_x$ and $\theta_y$ are azimuth angles of vector **r**'s projections on the *xz*-plane and *yz*-plane, respectively. $k_0$ is the wave vector in free space. $\varphi_{p,n_x n_y}$ is the entire phase of $PEU_{n_x n_y}$. $A_{x,n_x n_y}$ and $A_{y,n_x n_y}$ are modulated amplitudes of lights of *x*-polarization and *y*-polarization, respectively. $\varphi_{pol,n_x n_y}$ is the phase difference between two orthogonal polarization components. Considering each PEU has similar profile of emitted light in the far-field, the radiation vector can be approximated by $F_{n_x n_y}(\theta_x, \theta_y) \approx F(\theta_x, \theta_y)$. Furthermore, the denominator in Eq. S1 can be rewritten as $|\mathbf{r} - \mathbf{u}_{n_x n_y}| \approx |\mathbf{r}|$, while the phase factor in Eq. S1 can be approximated to a more precise form: $|\mathbf{r} - \mathbf{u}_{n_x n_y}| \approx |\mathbf{r}| - \mathbf{e}_r \cdot \mathbf{u}_{n_x n_y}$, where

$\mathbf{e}_r = [\sin\theta\cos\varphi \quad \sin\theta\sin\varphi \quad \cos\varphi] \approx [\sin\theta_x \quad \sin\theta_y \quad \cos\varphi]$ is the unit vector along the direction of **r** and $\mathbf{u}_{n_x n_y} = [\Lambda_x n_x \quad \Lambda_y n_y \quad 0]^T$ is the position vector of $PEU_{n_x n_y}$. $\Lambda_x$ and $\Lambda_y$ are the periods of adjacent PEUs along *x*-axis and *y*-axis, respectively. Using these approximations and assuming $\Lambda_x = \Lambda_y = \Lambda$, $N_x = N_y = N$, Eq. S1 can be simplified as

$$\begin{bmatrix} E_{x,PEA}(\mathbf{r}) \\ E_{y,PEA}(\mathbf{r}) \end{bmatrix} = F(\theta_x,\theta_y)\frac{e^{-jk_0|\mathbf{r}|}}{|\mathbf{r}|}\sum_{n_x=0}^{N-1}\sum_{n_y=0}^{N-1} e^{j\varphi_{p,n_xn_y}} \begin{bmatrix} A_{x,n_xn_y} \\ A_{y,n_xn_y} e^{j\varphi_{pol,n_xn_y}} \end{bmatrix} e^{j\Lambda k_0(n_x\sin\theta_x + n_y\sin\theta_y)}. \quad \text{(S2)}$$

In this article, the proposed optical lattice generator mainly utilizes part of the PEUs in the PEA, which are assembled in a square ring. Thus, the theory model of PEA can be further simplified, as shown in Fig. S4b. For ease of calculation, the PEUs in a square ring are divided into four parts, marked as $T_1$, $T_2$, $T_3$, and $T_4$, respectively. Therefore, Eq. S2 can be rewritten by

$$\begin{bmatrix} E_{x,PEA}(\mathbf{r}) \\ E_{y,PEA}(\mathbf{r}) \end{bmatrix} = F(\theta_x,\theta_y)\frac{e^{-jk_0|\mathbf{r}|}}{|\mathbf{r}|}(T_1+T_2+T_3+T_4), \quad \text{(S3)}$$

$$\begin{aligned}
T_1 &= \sum_{n_x=0}^{N-2} e^{j\varphi_{p,n_x0}} \begin{bmatrix} A_{x,n_x0} \\ A_{y,n_x0} e^{j\varphi_{pol,n_x0}} \end{bmatrix} e^{j\Lambda k_0 n_x \sin\theta_x} \\
T_2 &= \sum_{n_y=0}^{N-2} e^{j\varphi_{p,N-1n_y}} \begin{bmatrix} A_{x,N-1n_y} \\ A_{y,N-1n_y} e^{j\varphi_{pol,N-1n_y}} \end{bmatrix} e^{j\Lambda k_0 n_y \sin\theta_y} e^{j\Lambda k_0 (N-1)\sin\theta_x} \\
T_3 &= \sum_{n_x=N-1}^{1} e^{j\varphi_{p,n_xN-1}} \begin{bmatrix} A_{x,n_xN-1} \\ A_{y,n_xN-1} e^{j\varphi_{pol,n_xN-1}} \end{bmatrix} e^{j\Lambda k_0 n_x \sin\theta_x} e^{j\Lambda k_0 (N-1)\sin\theta_y} \\
T_4 &= \sum_{n_y=N-1}^{1} e^{j\varphi_{p,0n_y}} \begin{bmatrix} A_{x,0n_y} \\ A_{y,0n_y} e^{j\varphi_{pol,0n_y}} \end{bmatrix} e^{j\Lambda k_0 n_y \sin\theta_y}
\end{aligned} . \quad \text{(S4)}$$

For optical vortex (OV) lattice generation of arbitrary polarization state, $A_{x,n_xn_y}$, $A_{y,n_xn_y}$, and $\varphi_{pol,n_xn_y}$ are constant: $A_{x,n_xn_y} = A_x$, $A_{y,n_xn_y} = A_y$, $\varphi_{pol,n_xn_y} = \varphi_{pol}$. And the phase difference between adjacent PEUs in a square ring has fixed value of $\pi l / 2(N-1)$, where $l$ is the order of OV. Using Eq. S3 and Eq. S4, the Jones vector of the far-field light emitted from the PEA can be described as

$$\begin{bmatrix} E_{x,PEA}(\mathbf{r}) \\ E_{y,PEA}(\mathbf{r}) \end{bmatrix}_{OV} = \begin{bmatrix} A_x \\ A_y e^{j\varphi_{pol}} \end{bmatrix} F(\theta_x,\theta_y)\frac{e^{-jk_0|\mathbf{r}|}}{|\mathbf{r}|}(T_1+T_2+T_3+T_4), \quad \text{(S5)}$$

$$T_1=\sum_{n_x=0}^{N-2} e^{j\frac{\pi l}{2(N-1)}n_x} e^{j\Lambda k_0 n_x \sin\theta_x} = \frac{1-e^{j(N-1)\left[\frac{\pi l}{2(N-1)}+\Lambda k_0 \sin\theta_x\right]}}{1-e^{j\left[\frac{\pi l}{2(N-1)}+\Lambda k_0 \sin\theta_x\right]}}$$

$$T_2=\sum_{n_y=0}^{N-2} e^{j\frac{\pi l}{2}\left(1+\frac{n_y}{N-1}\right)} e^{j\Lambda k_0 n_y \sin\theta_y} e^{j\Lambda k_0 (N-1)\sin\theta_x} = \frac{1-e^{j(N-1)\left[\frac{\pi l}{2(N-1)}+\Lambda k_0 \sin\theta_y\right]}}{1-e^{j\left[\frac{\pi l}{2(N-1)}+\Lambda k_0 \sin\theta_y\right]}} e^{j\Lambda k_0 (N-1)\sin\theta_x} e^{j\frac{\pi}{2}l}$$

$$T_3=\sum_{n_x=N-1}^{1} e^{j\frac{\pi l}{2}\left(3-\frac{n_x}{N-1}\right)} e^{j\Lambda k_0 n_x \sin\theta_x} e^{j\Lambda k_0 (N-1)\sin\theta_y} = \frac{1-e^{j(N-1)\left[\frac{\pi l}{2(N-1)}-\Lambda k_0 \sin\theta_x\right]}}{1-e^{j\left[\frac{\pi l}{2(N-1)}-\Lambda k_0 \sin\theta_x\right]}} e^{j\Lambda k_0 (N-1)\sin\theta_x} e^{j\Lambda k_0 (N-1)\sin\theta_y} e^{j\pi l}$$

$$T_4=\sum_{n_y=N-1}^{1} e^{j\frac{\pi l}{2}\left(4-\frac{n_y}{N-1}\right)} e^{j\Lambda k_0 n_y \sin\theta_y} = \frac{1-e^{j(N-1)\left[\frac{\pi l}{2(N-1)}-\Lambda k_0 \sin\theta_y\right]}}{1-e^{j\left[\frac{\pi l}{2(N-1)}-\Lambda k_0 \sin\theta_y\right]}} e^{j\Lambda k_0 (N-1)\sin\theta_y} e^{j\frac{3\pi}{2}l}$$

. (S6)

When azimuth angle $\theta_x$ and $\theta_y$ satisfy the following conditions

$$\begin{aligned} \Lambda k_0 \sin\theta_x \approx \Lambda k_0 \theta_x = 2p\pi + \Lambda k_0 \Delta\theta \cos\varphi \quad (p=0,\pm1,\pm2,...) \\ \Lambda k_0 \sin\theta_y \approx \Lambda k_0 \theta_y = 2q\pi + \Lambda k_0 \Delta\theta \sin\varphi \quad (q=0,\pm1,\pm2,...) \end{aligned}, \quad \text{(S7)}$$

where $\Delta\theta$ is a slight offset to the azimuthal angle $\theta_x = 2p\pi/\Lambda k_0 = p\lambda_0/\Lambda$, $\theta_y = 2q\pi/\Lambda k_0 = q\lambda_0/\Lambda$, Eq. S6 can be changed to:

$$T_1=\frac{1-e^{j(N-1)\left[\frac{\pi l}{2(N-1)}+\Lambda k_0 \Delta\theta\cos\varphi\right]}}{1-e^{j\left[\frac{\pi l}{2(N-1)}+\Lambda k_0 \Delta\theta\cos\varphi\right]}} e^{j\Lambda k_0 (N-1)\Delta\theta\frac{1}{2}(\cos\varphi+\sin\varphi)} e^{j\Lambda k_0 (N-1)\Delta\theta\frac{1}{2}(-\cos\varphi-\sin\varphi)}$$

$$T_2=\frac{1-e^{j(N-1)\left[\frac{\pi l}{2(N-1)}+\Lambda k_0 \Delta\theta\sin\varphi\right]}}{1-e^{j\left[\frac{\pi l}{2(N-1)}+\Lambda k_0 \Delta\theta\sin\varphi\right]}} e^{j\Lambda k_0 (N-1)\Delta\theta\frac{1}{2}(\cos\varphi+\sin\varphi)} e^{j\Lambda k_0 (N-1)\Delta\theta\frac{1}{2}(\cos\varphi-\sin\varphi)} e^{j\frac{\pi}{2}l}$$

$$T_3=\frac{1-e^{j(N-1)\left[\frac{\pi l}{2(N-1)}-\Lambda k_0 \Delta\theta\cos\varphi\right]}}{1-e^{j\left[\frac{\pi l}{2(N-1)}-\Lambda k_0 \Delta\theta\cos\varphi\right]}} e^{j\Lambda k_0 (N-1)\Delta\theta\frac{1}{2}(\cos\varphi+\sin\varphi)} e^{j\Lambda k_0 (N-1)\Delta\theta\frac{1}{2}(\cos\varphi+\sin\varphi)} e^{j\pi l}$$

$$T_4=\underbrace{\frac{1-e^{j(N-1)\left[\frac{\pi l}{2(N-1)}-\Lambda k_0 \Delta\theta\sin\varphi\right]}}{1-e^{j\left[\frac{\pi l}{2(N-1)}-\Lambda k_0 \Delta\theta\sin\varphi\right]}}}_{1} \underbrace{e^{j\Lambda k_0 (N-1)\Delta\theta\frac{1}{2}(\cos\varphi+\sin\varphi)}}_{2} \underbrace{e^{j\Lambda k_0 (N-1)\Delta\theta\frac{1}{2}(-\cos\varphi+\sin\varphi)} e^{j\frac{3\pi}{2}l}}_{3}$$

. (S8)

One can see from Eq. S8 that it is composed of three parts: the first one is the complex amplitude term; the second one is a fixed phase factor; the third one can be seen as a uniform circular array of 4 elements with constant phase difference of $\pi l/2$. When the slight offset $\Delta\theta$ tends to 0, the complex amplitude term can be approximated as a constant one and Eq. S8 is expressed as

$$T_1 = \frac{\sin\left(\frac{\pi l}{4}\right)}{\sin\left[\frac{\pi l}{4(N-1)}\right]} e^{j\frac{\pi l(N-2)}{4(N-1)}} e^{j\Lambda k_0(N-1)\Delta\theta\frac{1}{2}(\cos\varphi+\sin\varphi)} e^{j\Lambda k_0(N-1)\Delta\theta\frac{1}{2}(-\cos\varphi-\sin\varphi)}$$

$$T_2 = \frac{\sin\left(\frac{\pi l}{4}\right)}{\sin\left[\frac{\pi l}{4(N-1)}\right]} e^{j\frac{\pi l(N-2)}{4(N-1)}} e^{j\Lambda k_0(N-1)\Delta\theta\frac{1}{2}(\cos\varphi+\sin\varphi)} e^{j\Lambda k_0(N-1)\Delta\theta\frac{1}{2}(\cos\varphi-\sin\varphi)} e^{j\frac{\pi}{2}l}$$

$$T_3 = \frac{\sin\left(\frac{\pi l}{4}\right)}{\sin\left[\frac{\pi l}{4(N-1)}\right]} e^{j\frac{\pi l(N-2)}{4(N-1)}} e^{j\Lambda k_0(N-1)\Delta\theta\frac{1}{2}(\cos\varphi+\sin\varphi)} e^{j\Lambda k_0(N-1)\Delta\theta\frac{1}{2}(\cos\varphi+\sin\varphi)} e^{j\pi l}$$

$$T_4 = \frac{\sin\left(\frac{\pi l}{4}\right)}{\sin\left[\frac{\pi l}{4(N-1)}\right]} e^{j\frac{\pi l(N-2)}{4(N-1)}} e^{j\Lambda k_0(N-1)\Delta\theta\frac{1}{2}(\cos\varphi+\sin\varphi)} e^{j\Lambda k_0(N-1)\Delta\theta\frac{1}{2}(-\cos\varphi+\sin\varphi)} e^{j\frac{3\pi}{2}l} \quad . \qquad \text{(S9)}$$

Substituting Eq. S9 into Eq. S5, the Jones vector of the emitted light can be expressed as[2]

$$\begin{aligned}
\begin{bmatrix} E_{x,PEA}(\mathbf{r}) \\ E_{y,PEA}(\mathbf{r}) \end{bmatrix}_{OV} &= \begin{bmatrix} A_x \\ A_y e^{j\varphi_{pol}} \end{bmatrix} F(\theta_x,\theta_y) \frac{e^{-jk_0|\mathbf{r}|}}{|\mathbf{r}|} \cdot \frac{\sin\left(\frac{\pi l}{4}\right)}{\sin\left[\frac{\pi l}{4(N-1)}\right]} e^{j\frac{\pi l(N-2)}{4(N-1)}} \cdot \\
&\quad e^{j\Lambda k_0(N-1)\Delta\theta\frac{1}{\sqrt{2}}\sin\left(\varphi+\frac{\pi}{4}\right)} e^{j\frac{3}{4}\pi l} \cdot \sum_{s=1}^{4} e^{j\Lambda k_0(N-1)\Delta\theta\frac{1}{\sqrt{2}}\cos\left(\varphi-\frac{2s-1}{4}\pi\right)} e^{j\frac{2s-1}{4}\pi l} \quad , \text{(S10)} \\
&= \begin{bmatrix} A_x \\ A_y e^{j\varphi_{pol}} \end{bmatrix} F(\theta_x,\theta_y) \frac{e^{-jk_0|\mathbf{r}|}}{|\mathbf{r}|} \cdot \frac{\sin\left(\frac{\pi l}{4}\right)}{\sin\left[\frac{\pi l}{4(N-1)}\right]} e^{j\frac{\pi l(4N-5)}{4(N-1)}} \cdot \\
&\quad e^{j\Lambda k_0(N-1)\Delta\theta\frac{1}{\sqrt{2}}\sin\left(\varphi+\frac{\pi}{4}\right)} \cdot 4 j^{|l|} J_{|l|}\left(\frac{1}{\sqrt{2}}\Lambda k_0(N-1)\Delta\theta\right) e^{jl\varphi}
\end{aligned}$$

where $J_{|l|}$ represents the |*l*|-order Bessel functions of the first kind. Combing Eq. S7 and Eq. S7, one can indicate that the helical phase factor $e^{jl\varphi}$ simultaneously appears at the positions of $\theta_x = p\lambda_0/\Lambda$, $\theta_y = q\lambda_0/\Lambda$ ($p,q = 0,\pm1,\pm2,...$) in the far-field, meaning the generation of OV lattice with rectangular net in the far-field with azimuth angle period of $\lambda_0/\Lambda$.

For the generation of cylindrical vector beam (CVB) lattice with circularly symmetric linearly-polarized (LP) distribution, $\varphi_{p,n_xn_y}$, and $\varphi_{pol,n_xn_y}$ are constant: $\varphi_{p,n_xn_y} = \varphi_{pol,n_xn_y} = 0$. Equation S4 can be rewritten as

$$
\begin{aligned}
T_1 &= \sum_{n_x=0}^{N-2} \begin{bmatrix} \cos\left(\varphi_0 + \dfrac{\pi l_p}{2(N-1)} n_x\right) \\ \sin\left(\varphi_0 + \dfrac{\pi l_p}{2(N-1)} n_x\right) \end{bmatrix} e^{j\Lambda k_0 n_x \sin\theta_x} \\
T_2 &= \sum_{n_y=0}^{N-2} \begin{bmatrix} \cos\left(\varphi_0 + \dfrac{\pi l_p}{2}\left(1+\dfrac{n_y}{N-1}\right)\right) \\ \sin\left(\varphi_0 + \dfrac{\pi l_p}{2}\left(1+\dfrac{n_y}{N-1}\right)\right) \end{bmatrix} e^{j\Lambda k_0 n_y \sin\theta_y} e^{j\Lambda k_0 (N-1)\sin\theta_x} \\
T_3 &= \sum_{n_x=N-1}^{1} \begin{bmatrix} \cos\left(\varphi_0 + \dfrac{\pi l_p}{2}\left(3-\dfrac{n_x}{N-1}\right)\right) \\ \sin\left(\varphi_0 + \dfrac{\pi l_p}{2}\left(3-\dfrac{n_x}{N-1}\right)\right) \end{bmatrix} e^{j\Lambda k_0 n_x \sin\theta_x} e^{j\Lambda k_0 (N-1)\sin\theta_y} \\
T_4 &= \sum_{n_y=N-1}^{1} \begin{bmatrix} \cos\left(\varphi_0 + \dfrac{\pi l_p}{2}\left(4-\dfrac{n_y}{N-1}\right)\right) \\ \sin\left(\varphi_0 + \dfrac{\pi l_p}{2}\left(4-\dfrac{n_y}{N-1}\right)\right) \end{bmatrix} e^{j\Lambda k_0 n_y \sin\theta_y}
\end{aligned}
\quad , \qquad \text{(S11)}
$$

where $\varphi_0$ and $l_p$ are the polarization orientation angle and the order of polarization state of CVB, respectively. Using the similar analysis method in Eq. S7-S10, the Jones vector of the far-field light emitted from the PEA can be expressed as

$$\begin{bmatrix} E_{x,PEA}(\mathbf{r}) \\ E_{y,PEA}(\mathbf{r}) \end{bmatrix}_{CVB} = F(\theta_x,\theta_y)\frac{e^{-jk_0|\mathbf{r}|}}{|\mathbf{r}|}\cdot\frac{\sin\left(\frac{\pi l}{4}\right)}{\sin\left[\frac{\pi l}{4(N-1)}\right]}\cdot e^{j\Lambda k_0(N-1)\Delta\theta\frac{1}{\sqrt{2}}\sin\left(\varphi+\frac{\pi}{4}\right)}\cdot$$
$$4j^{|l_p|}J_{|l_p|}\left(\frac{1}{\sqrt{2}}\Lambda k_0(N-1)\Delta\theta\right)\cdot\begin{bmatrix}\cos\left(\varphi_0+l_p\left(\varphi+\frac{4N-5}{4N-4}\pi\right)\right)\\ \sin\left(\varphi_0+l_p\left(\varphi+\frac{4N-5}{4N-4}\pi\right)\right)\end{bmatrix} \quad \text{(S12)}$$

From Eq. S7 and Eq. S12, it is clear that multiple CVBs locate at the positions of $\theta_x = p\lambda_0/\Lambda$, $\theta_y = q\lambda_0/\Lambda$ ($p,q = 0,\pm1,\pm2,\ldots$) in the far field with azimuth angle period of $\lambda_0/\Lambda$, proving the generation of CVB lattice.

For the generation of vector vortex beam (VVB) lattices with circularly symmetric LP distribution, $\varphi_{pol,n_xn_y}$ meets the condition of $\varphi_{pol,n_xn_y}=0$. Equation S4 can be expressed in a new form

$$T_1=\sum_{n_x=0}^{N-2} e^{j\frac{\pi l_o}{2(N-1)}n_x}\begin{bmatrix}\cos\left(\varphi_0+\frac{\pi l_p}{2(N-1)}n_x\right)\\ \sin\left(\varphi_0+\frac{\pi l_p}{2(N-1)}n_x\right)\end{bmatrix}e^{j\Lambda k_0 n_x\sin\theta_x}$$
$$T_2=\sum_{n_y=0}^{N-2} e^{j\frac{\pi l_o}{2}\left(1+\frac{n_y}{N-1}\right)}\begin{bmatrix}\cos\left(\varphi_0+\frac{\pi l_p}{2}\left(1+\frac{n_y}{N-1}\right)\right)\\ \sin\left(\varphi_0+\frac{\pi l_p}{2}\left(1+\frac{n_y}{N-1}\right)\right)\end{bmatrix}e^{j\Lambda k_0 n_y\sin\theta_y}e^{j\Lambda k_0(N-1)\sin\theta_x}$$
$$T_3=\sum_{n_x=N-1}^{1} e^{j\frac{\pi l_o}{2}\left(3-\frac{n_y}{N-1}\right)}\begin{bmatrix}\cos\left(\varphi_0+\frac{\pi l_p}{2}\left(3-\frac{n_y}{N-1}\right)\right)\\ \sin\left(\varphi_0+\frac{\pi l_p}{2}\left(3-\frac{n_y}{N-1}\right)\right)\end{bmatrix}e^{j\Lambda k_0 n_x\sin\theta_x}e^{j\Lambda k_0(N-1)\sin\theta_y}$$
$$T_4=\sum_{n_y=N-1}^{1} e^{j\frac{\pi l_o}{2}\left(4-\frac{n_y}{N-1}\right)}\begin{bmatrix}\cos\left(\varphi_0+\frac{\pi l_p}{2}\left(4-\frac{n_y}{N-1}\right)\right)\\ \sin\left(\varphi_0+\frac{\pi l_p}{2}\left(4-\frac{n_y}{N-1}\right)\right)\end{bmatrix}e^{j\Lambda k_0 n_y\sin\theta_y} \quad , \text{(S13)}$$

where $l_o$ and $l_p$ are the order of OV and polarization state of VVB, respectively. $\varphi_0$ is the polarization orientation angle. Utilizing Eq. S3, Eq. S7, and Eq. S13, the Jones vector of the far-field light emitted from the PEA can be expressed as

$$\begin{bmatrix} E_{x,PEA}(\mathbf{r}) \\ E_{y,PEA}(\mathbf{r}) \end{bmatrix}_{VVB} = F(\theta_x,\theta_y)\frac{e^{-jk_0|\mathbf{r}|}}{|\mathbf{r}|}\cdot\frac{\sin\left(\frac{\pi l}{4}\right)}{\sin\left[\frac{\pi l}{4(N-1)}\right]}\cdot e^{j\Lambda k_0(N-1)\Delta\theta\frac{1}{\sqrt{2}}\sin\left(\varphi+\frac{\pi}{4}\right)}\cdot$$
$$4j^{|l_p|}J_{|l_p|}\left(\frac{1}{\sqrt{2}}\Lambda k_0(N-1)\Delta\theta\right)\cdot\begin{bmatrix}\cos\left(\varphi_0+l_p\left(\varphi+\frac{4N-5}{4N-4}\pi\right)\right)\\ \sin\left(\varphi_0+l_p\left(\varphi+\frac{4N-5}{4N-4}\pi\right)\right)\end{bmatrix}\cdot e^{jl_o\left(\varphi+\frac{4N-5}{4N-4}\pi\right)} \quad .(S14)$$

From Eq. S7 and Eq. S14, one can indicate that multiple VVBs exist at the points of $\theta_x = 2p\pi/\Lambda k_0$ , $\theta_y = 2q\pi/\Lambda k_0$ ( $p,q = 0,\pm1,\pm2,...$ ) in the far field, proving the generation of VVB lattice.

Combining Eq. S10, Eq. S12, and Eq. S14, the azimuth angle period of various optical lattice can be described as

$$\Lambda_\theta = \frac{2\pi}{\Lambda k_0} = \frac{\lambda_0}{\Lambda}, \quad (S15)$$

where $\lambda_0$ is the wavelength of light in free space. Equation S15 reveals that the azimuth angle period in the far-field is decided by the wavelength and the periods of adjacent PEUs.

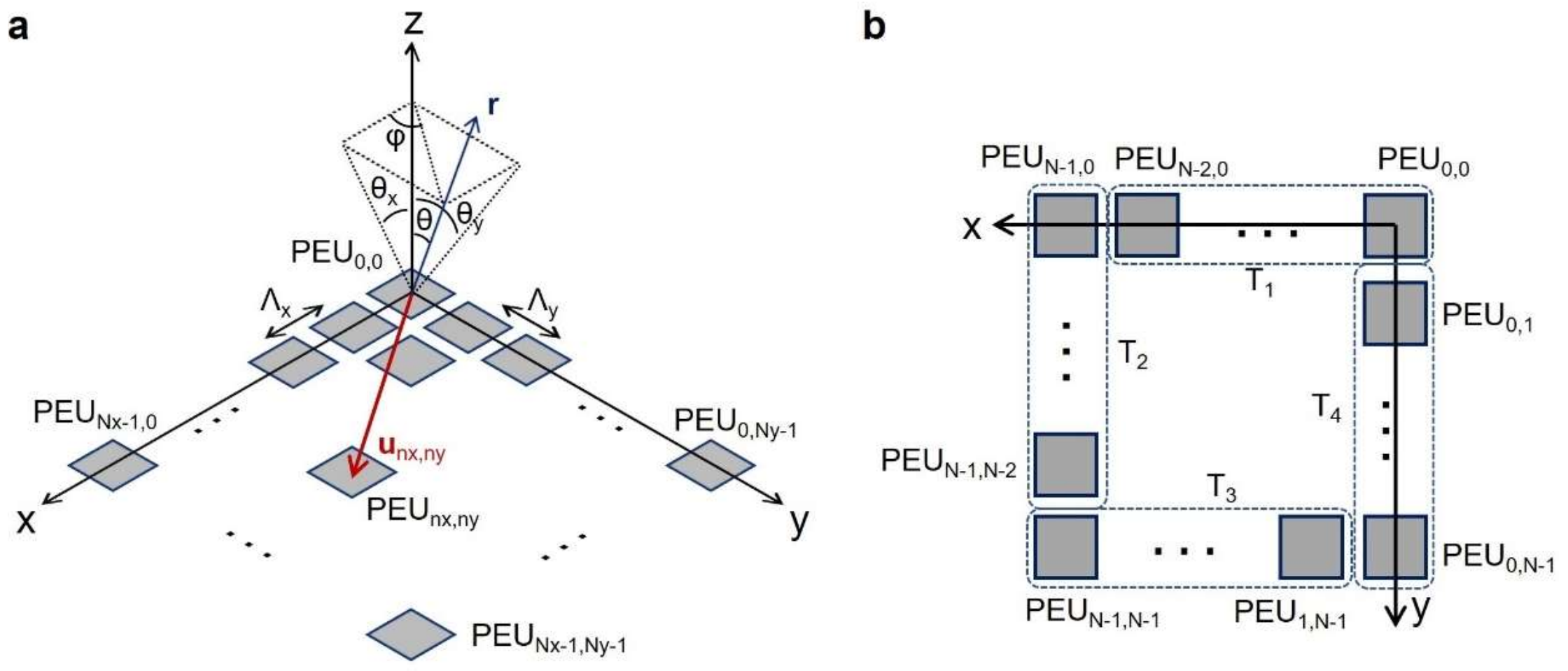

**Figure S4 | Schematic of PEA**. **a**, Light diffraction model of an ideal rectangular array. **b**, A kind of rectangular array where PEUs are assembled in a square ring.

## 5. Modulation speed

The modulation speed of the fabricated chip is dependent on the response speed of the thermal-optic microheaters. The thermal-optic effect of TiN microheaters is used to tune the optical path, so the response time is usually in the microsecond range. As shown in Fig. S5a, a 5-kHz square-wave voltage signal with a 50% duty cycle is applied to a heater, and the temporal waveform is detected by a high-speed detector. The results in Fig. S4a show that the rising time (10%~90%) and falling time (90%~10%) of electrical signal upon thermal tuning is about 15 μs. More temporal responses of square-wave voltage signal with different frequency and duty circle are also depicted in Figs. S5b-S5f.

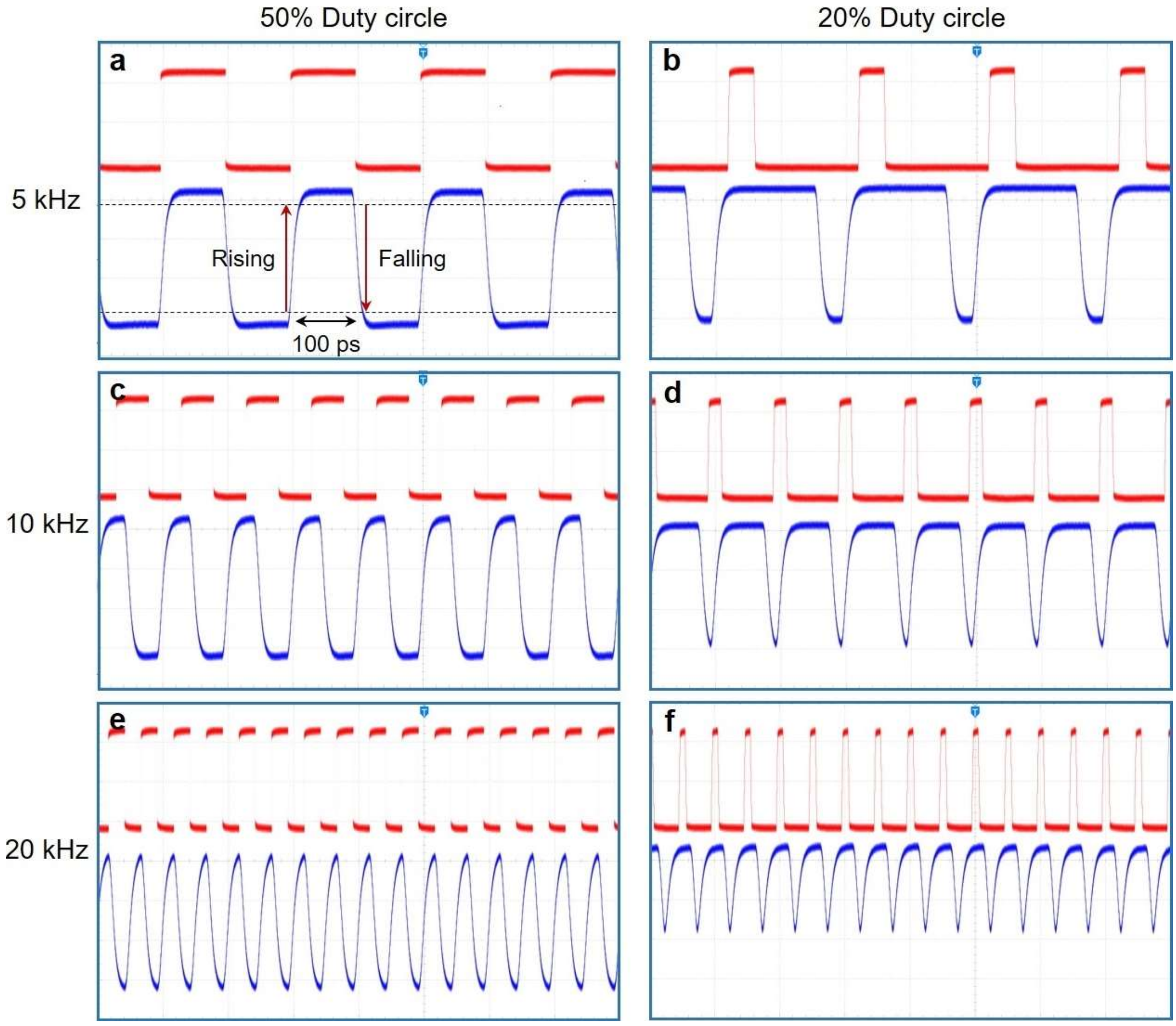

**Figure S5 | Experimentally measured time domain responses for the thermal microheaters**.

## 6. Diffraction processes for OV lattices

To display the evolution process of the emitted light from the PEA, we simulate the light field distributions at different propagation distances. As shown in the first row of Figs. S6a and S6b, to generate OV lattices with topological charge of -1 and +2, 8 PEUs in a square with uniform intensities and gradual phase change are used in the simulation. At the distance of $z$ = 0.1 mm, the Gaussian beams emitted from PEA are observed. With the increase of propagation distance from $z$ = 0.1 mm to $z$ = 0.6 m, the OV lattice can be produced in the far-field. Additionally, we investigate the impact of the intensity distribution of the PEA on the generated OV lattice. As shown in the second row of Figs. S6a and S6b, 8 PEUs in a square with nonuniform intensities can also generate OV lattices with topological charge of -1 and +2 in the far-field, where the four PEUs on the diagonal have half the intensity of the other ones. Compared the intensity profiles in the first and second row of Figs. S6a and b, one can indicate that the generated OV beams in the first row have more uniform intensity distribution characteristics in the azimuthal direction than that in the second row.

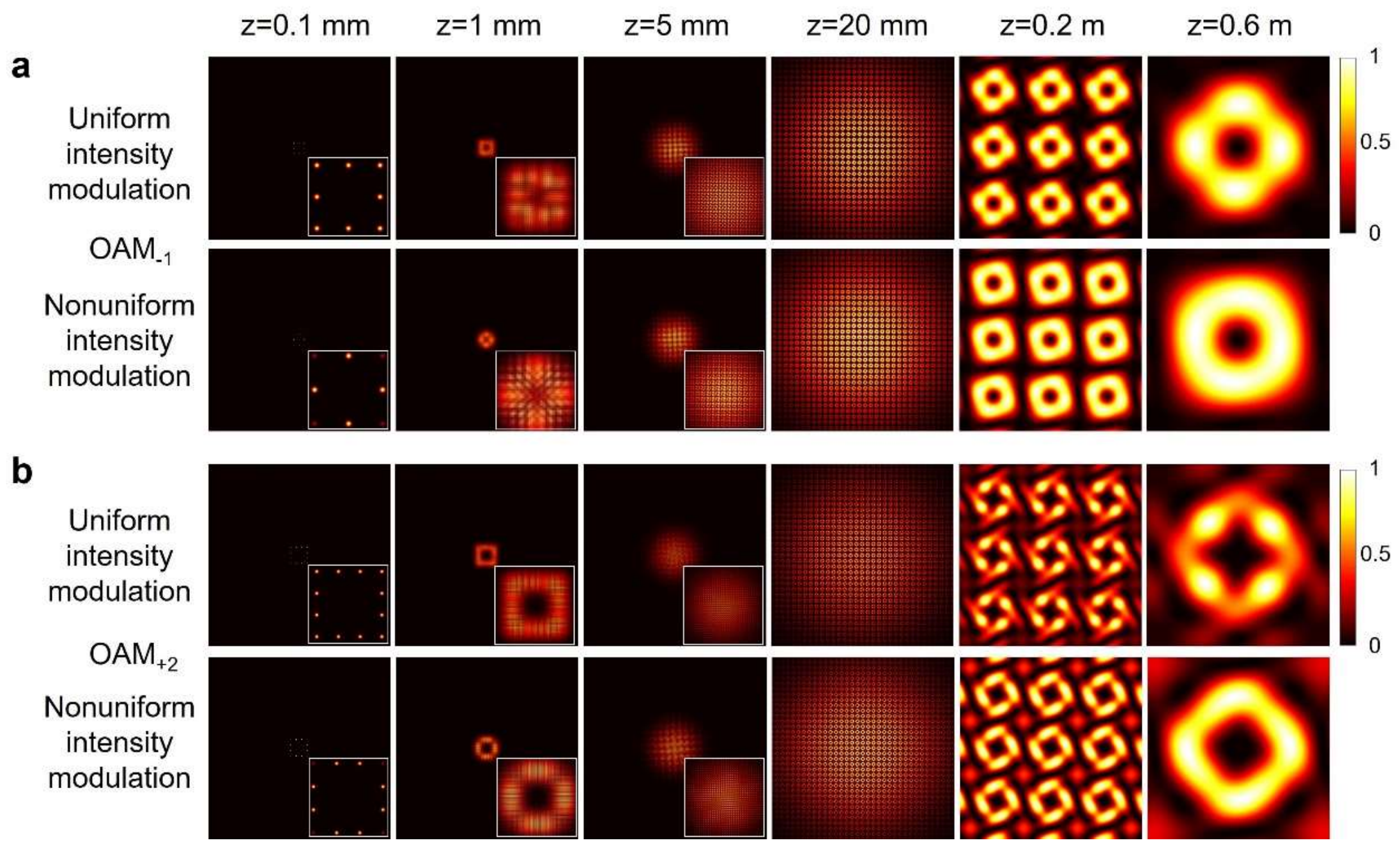

**Figure S6 | Simulated diffraction processes for OV lattices. a,** 8 PEUs in a square are utilized for the generation of OV lattice with $OAM_{-1}$. **b,** 12 PEUs in a square are utilized for the generation of OV lattice with $OAM_{+2}$.

## 7. Additional results for the generation of OV lattices

Here, 8 PEUs in a square with uniform intensities and gradual phase change are used are chosen for the simultaneous generation of *x*- and *y*-polarized OV lattices with $OAM_{-1}$. As shown in Fig. S7, the first column shows the intensity profiles of the generated OV lattice for both *x*- and *y*-polarization in the far-field at 1550 nm. The second column shows the zoom-in views, from which one can see the donut-shaped intensity profiles due to the phase singularity. The clockwise spiral and fork patterns for the collinear and tilt interferograms are also obtained in the third and fourth column, respectively, which verify the spiral phase structures.

The broadband characteristic of the generated OV lattice is further demonstrated in Figs. S8 and Figs. S9. We choose *x*-polarized OV lattice with $OAM_{-1}$ as one example, as shown in Fig. S8. The measured intensity profiles, the collinear and tilt interferograms at the wavelength of 1530, 1550 and 1570 nm are obtained, respectively. We then characterize the mode quality by focusing on the evaluation of the phase purity. Figure. S9 depicts the measured phase purities of the generated *x*-polarized beams with $OAM_{-1}$ in OV lattice as a function of the wavelength (covering almost whole C- and L-band) from 1528.3 nm to 1600.6 nm, where the calculated average phase purity is about 0.88.

The broadband characteristic of the generated OV lattice with higher topological charge number is also investigated. We use 12 PEUs in a square with uniform intensities and gradual phase change to generate OV lattice with $OAM_{\pm 2}$. As shown in Fig. S10, the measured intensity profiles, the collinear and tilt interferograms of OV lattice at the wavelength of 1530, 1550 and 1570 nm are presented. One can indicate that the collinear interferograms of *x*-polarized beams with $OAM_{\pm 2}$ in OV lattice have two arms but opposite spiral directions, proving the topological charge number. All the measured results verify the feasibility of broadband OV lattice generation based on the fabricated PEA.

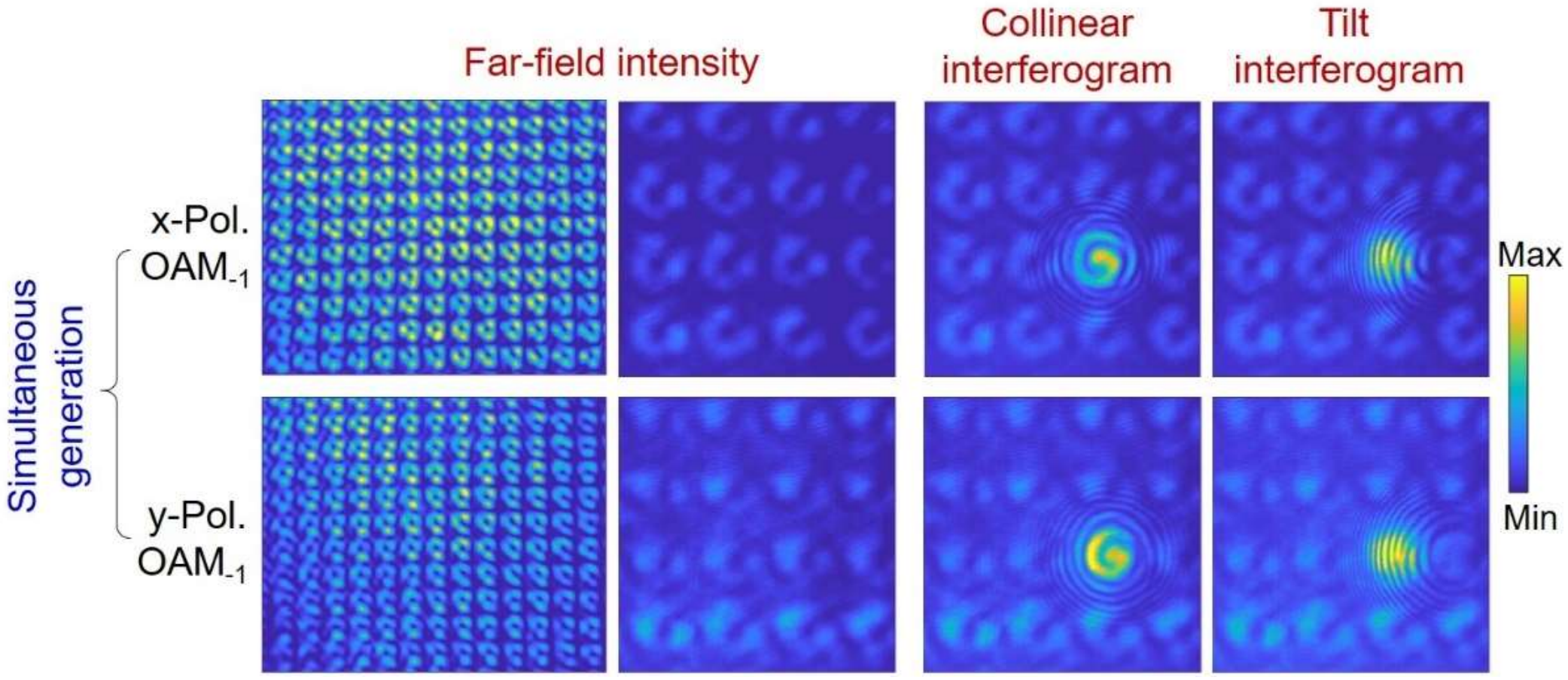

**Figure S7 | Experimentally measured results for the simultaneous generation of OV lattices with two polarization states**.

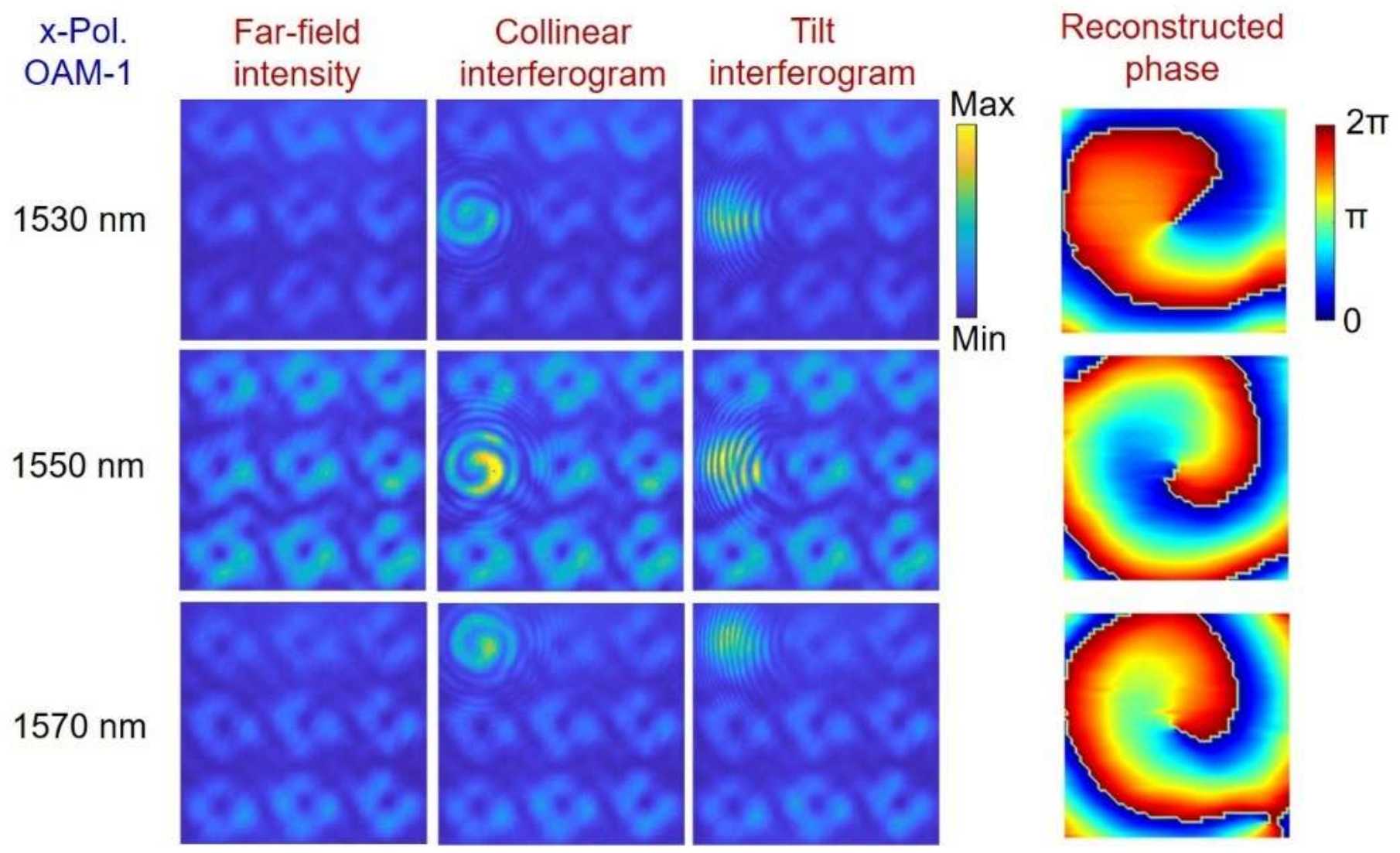

**Figure S8 | Experimentally measured results for broadband generation of *x*-polarized OV lattice with OAM$_{-1}$.**

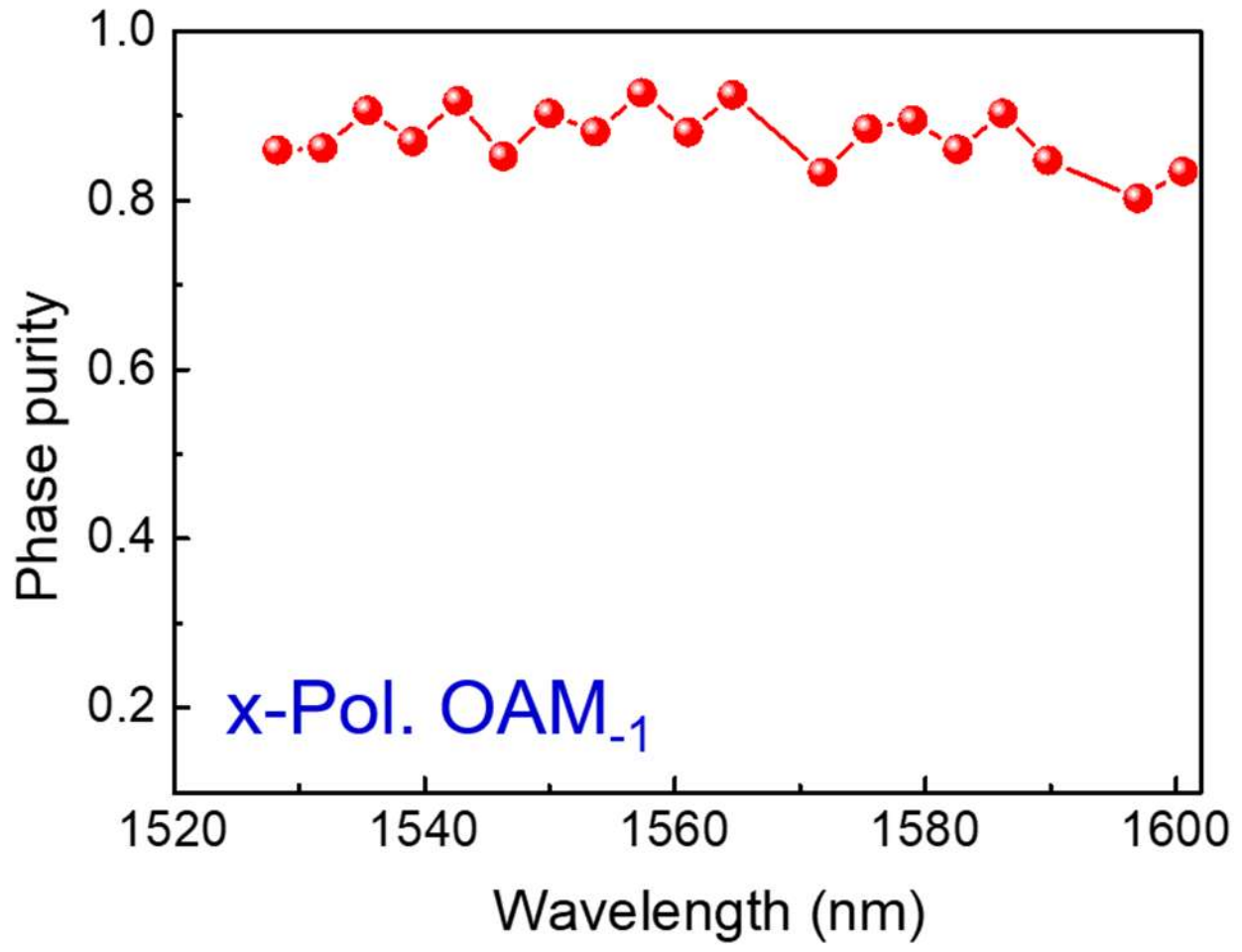

**Figure S9 | Calculated phase purity of *x*-polarized OV lattice with OAM$_{-1}$ versus wavelength.**

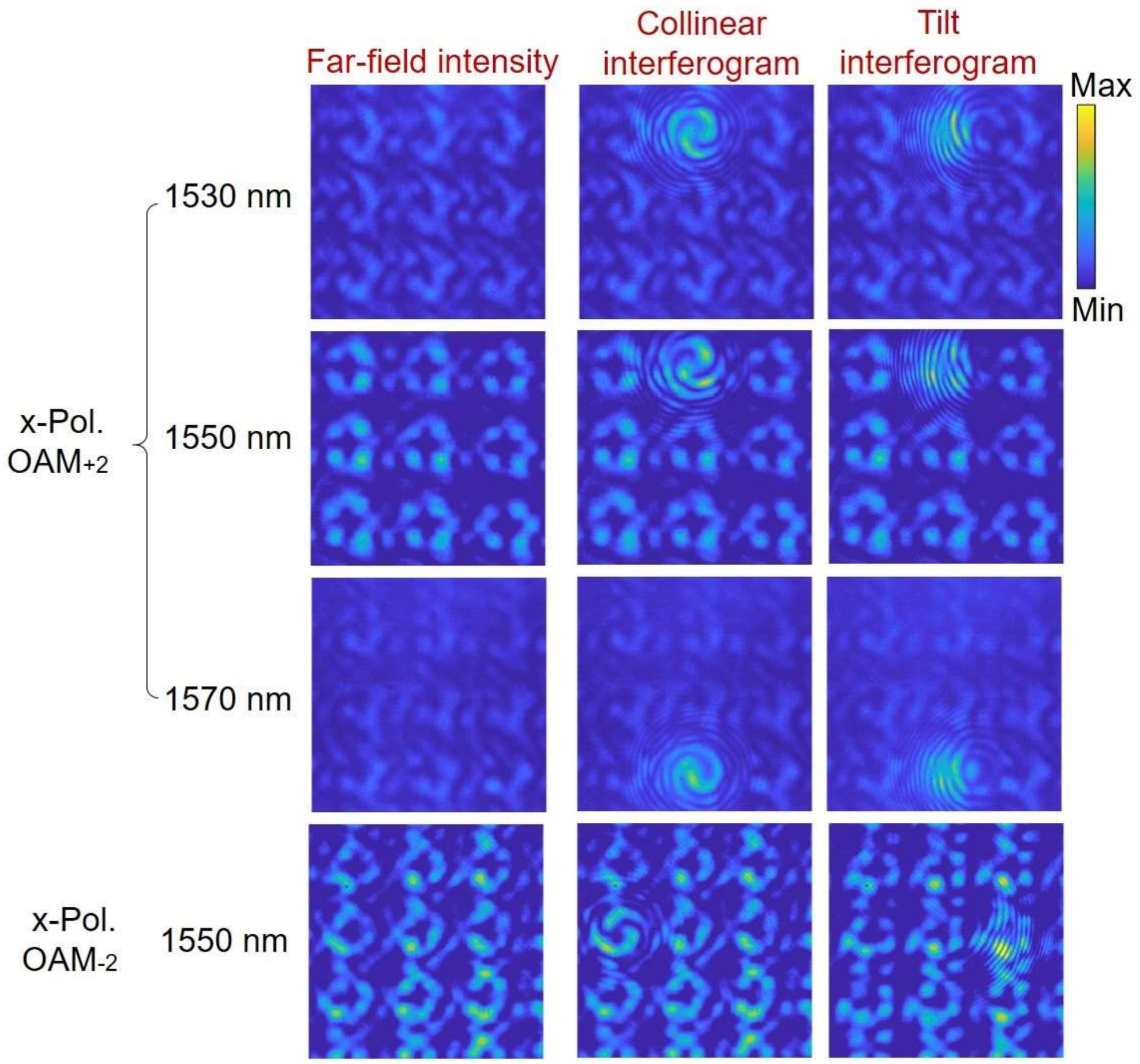

**Figure S10 | Experimentally measured results for broadband generation of *x*-polarized OV lattice with OAM$_{\pm 2}$.**

## 8. Additional results for the generation of CVB lattices

The generation of CVB lattice of other states using the designed PEA is also investigated, as shown in Fig. S11. Similar to Eq. 5, the Jones vector of CVBs with circularly symmetric LP distribution can be expressed in another form

$$E_{CVB}^{LP} = \begin{bmatrix} \cos\left(\varphi_0 + l_p \varphi\right) \\ -\sin\left(\varphi_0 + l_p \varphi\right) \end{bmatrix} = \frac{1}{\sqrt{2}}\left(e^{j\varphi_0} e^{jl_p\varphi} \mathbf{L} + e^{-j\varphi_0} e^{-jl_p\varphi} \mathbf{R}\right), \quad \text{(S15)}$$

where **L** and **R** are left-handed circularly polarized (LCP) and right-handed circularly polarized (RCP) components, respectively, whose Jones vectors are $\begin{bmatrix}1 & \pm j\end{bmatrix}^T / \sqrt{2}$ . $\varphi_0$ and $l_p$ are the polarization orientation angle and the order of polarization state of CVB, respectively. Figure S11a shows a Poincaré sphere with $l_p$ = +1, where the poles represent two OV beams with uniform amplitude but opposite helicity and polarization (LCP $OAM_{+1}$ at the north pole and RCP $OAM_{-1}$ at the south pole). Figures S11b-S11e respectively display the measured results of CVB lattice generation with $l_p$ = +1, $\varphi_0$ = 0°, 45°, 90°, and 135°. One can see that the donut-shaped intensity profiles of CVBs in the lattice become two-lobed after passing through the polarizer, where the orientation angle of dark fringes is relative to the polarization direction of polarizer, verified the polarization states of generated CVBs. Moreover, the broadband feature of CVB lattice generation is demonstrated. Figure S12 depicts the generated CVB lattice with $l_p$ = +1, $\varphi_0$ = 0° (composed of LCP $OAM_{-1}$ and RCP $OAM_{+1}$) at the wavelength of 1530, 1540, 1550, and 1560 nm, respectively. Polarizers are also applied to analyze the polarization distribution of generated optical lattice, where the two-lobed patterns rotate with the orientation change of the polarizer, confirming the polarization state of CVB lattice. The results indicate that the designed PEA can generate CVB lattice covering nearly 30 nm wavelength range from 1530 to 1560 nm.

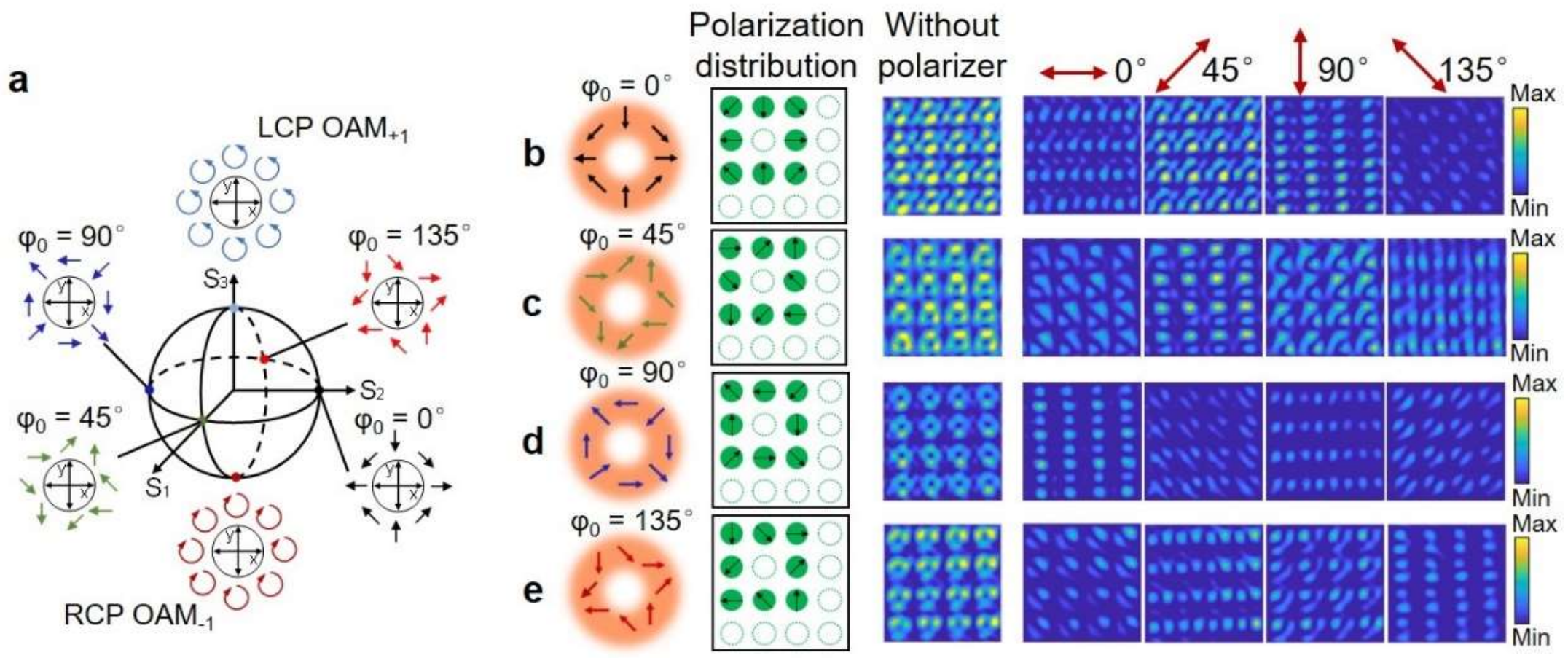

**Figure S11 | Experimentally measured results for CVB lattice generation. a,** A Poincaré sphere with $l_p$ = +1 (LCP $OAM_{+1}$ at the north pole and RCP $OAM_{-1}$ at the south pole). **b-e,** 8 PEUs in a square are used to generate CVB lattice with $\varphi_0$ = 0°, 45°, 90°, and 135°, respectively.

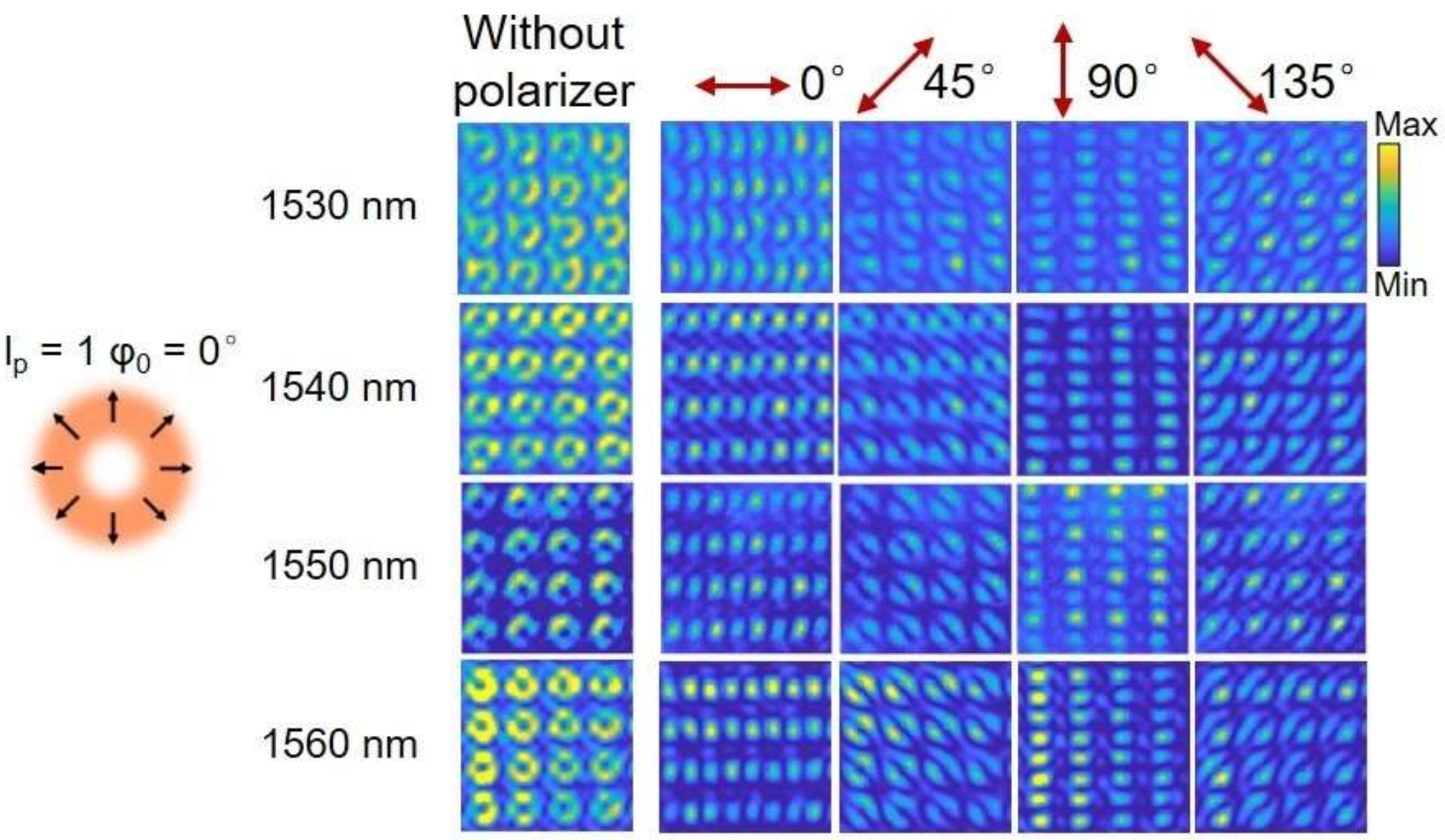

**Figure S12 | Experimentally measured results for broadband generation of CVB lattice with $l_p$ = +1 and $\varphi_0$ = 0°.**

## 9. Polarization reconstruction of generated CVB lattices

To obtain the polarization reconstruction of the generated CVB lattice, we use Stokes polarimetry where the Stokes parameters are measured via intensity measurements. The polarization distribution of an optical beam can be determined by the following four Stokes parameters

$$S_0(x,y)=I_0(x,y)+I_{90}(x,y)\,, \tag{S16}$$

$$S_1(x,y)=I_0(x,y)-I_{90}(x,y)\,, \tag{S17}$$

$$S_2(x,y)=I_{45}(x,y)-I_{135}(x,y)\,, \tag{S18}$$

$$S_3(x,y)=I_R(x,y)-I_L(x,y)\,, \tag{S19}$$

where $I_0(x, y)$, $I_{45}(x, y)$, $I_{90}(x, y)$, and $I_{135}(x, y)$ are the intensities of the optical beam passing through a linear polarizer oriented at 0°, 45°, 90° and 135°, respectively. $I_R(x, y)$ and $I_L(x, y)$ are the intensity of RCP and LCP components, respectively. By measuring the intensity components of $I_0(x, y)$, $I_{45}(x, y)$, $I_{90}(x, y)$, $I_{135}(x, y)$, $I_R(x, y)$, and $I_L(x, y)$, one can reconstruct the polarization distribution of an optical beam. Figure S13 shows the simulated and measured

CVB lattice with $l_p$ = +1, $\varphi_0$ = 0° (composed of LCP $OAM_{-1}$ and RCP $OAM_{+1}$). The recovered polarization distribution of generated CVB lattice is also displayed in Fig. S13 by utilizing the measured six intensity components. One can indicate that the reconstructed polarization profile has circularly symmetric LP distribution, which fits the simulated result well and verified the polarization state of generated CVB lattice.

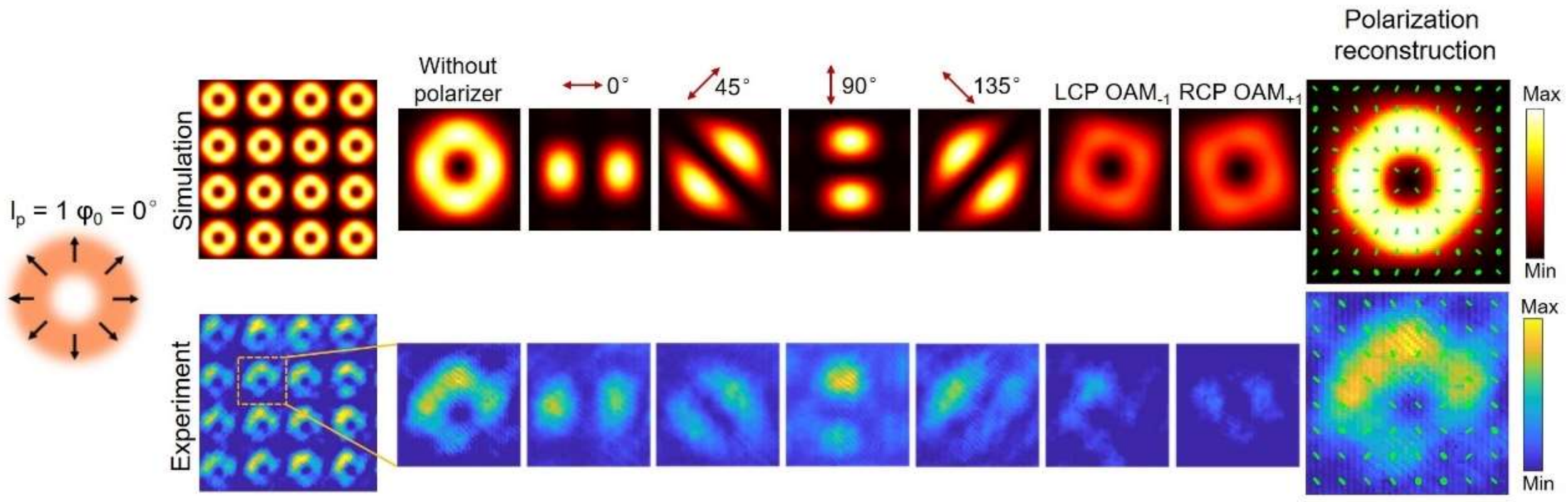

**Figure S13 | Polarization reconstruction of CVB lattice with $l_p$ = +1 and $\varphi_0$ = 0°.**

## References

1. Van Acoleyen, K., Nanophotonic beamsteering elements using silicon technology for wireless optical applications, (Ghent University, 2012).
2. Van Trees, H. L., Optimum array processing: Part IV of detection, estimation, and modulation theory (John Wiley & Sons, 2004).